\documentclass[aps,preprintnumbers,twocolumn,nofootinbib,notitlepage,superscriptaddress]{revtex4-2}

\usepackage{natbib}
\usepackage{graphicx}
\usepackage{dcolumn}
\usepackage{lmodern}
\usepackage{slashed}
\usepackage{xcolor}
\usepackage{mathtools,braket}
\usepackage{physics}
\usepackage{multirow}
\usepackage[colorlinks=true, pdfstartview=FitV, bookmarks=true,
bookmarksnumbered=true,breaklinks]{hyperref}
\usepackage{soul}
\usepackage{lipsum}  
\usepackage{color}
\usepackage{booktabs}
\definecolor{blue}{rgb}{0.0, 0.0, 1.0}
\definecolor{red}{rgb}{1.0, 0.0, 0.0}
\definecolor{royalblue}{rgb}{0.0, 0.14, 0.4}
\hypersetup{linkcolor=blue, citecolor=blue, urlcolor=blue}

\usepackage{newtxmath}
\usepackage{bm}

\usepackage{amsmath,amssymb}
\usepackage{mathrsfs}

\usepackage{hyperref}
\hypersetup{colorlinks=true,citecolor=blue,linkcolor=blue,urlcolor=blue}

\def\orcid#1{\kern .08em\href{https://orcid.org/#1}
{\includegraphics[keepaspectratio,width=0.7em]{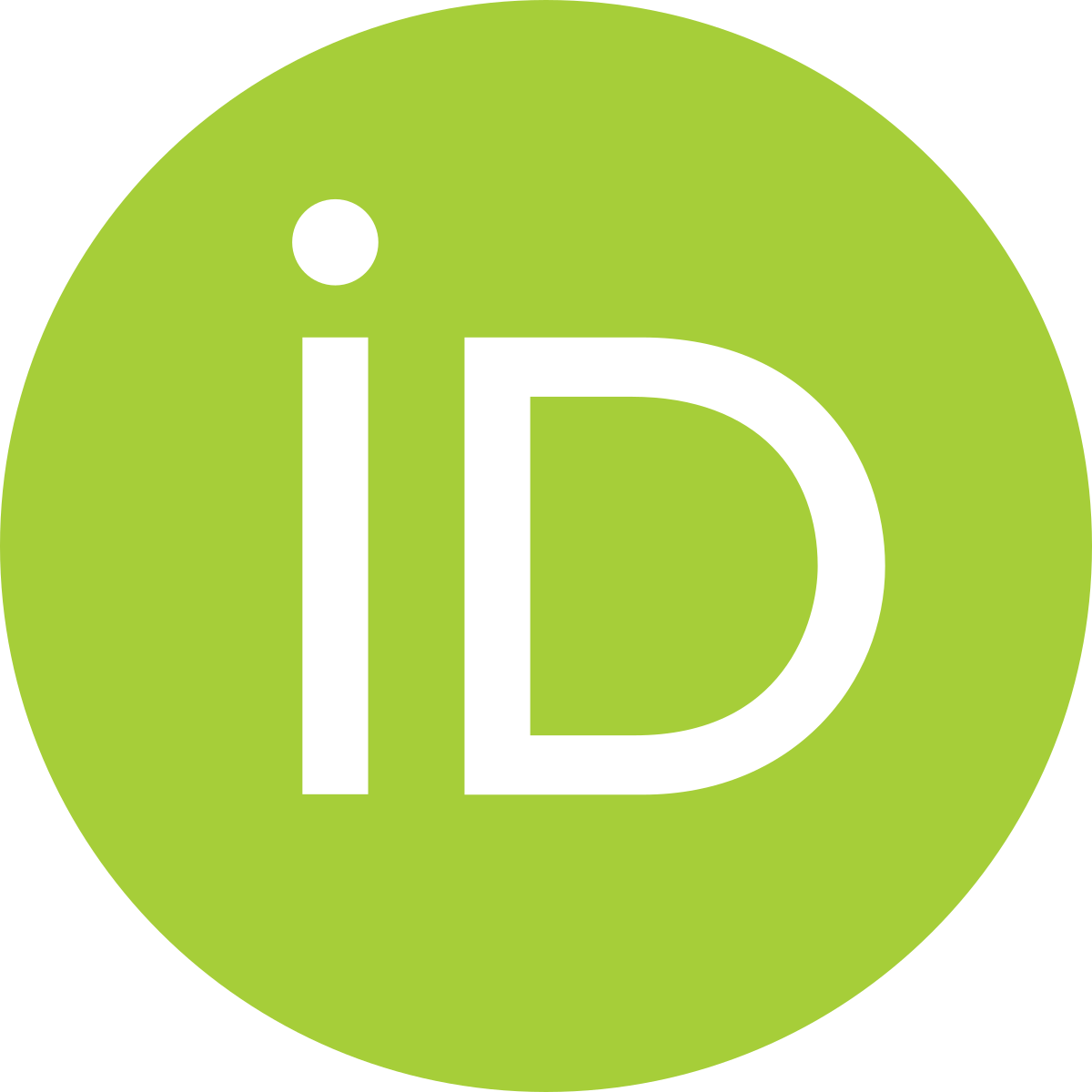}}}
\usepackage[mathlines]{lineno}

\begin{document}
\preprint{LFTC-26/05/114}
\title{Medium modification of pseudoscalar meson structure: \\Distribution amplitudes and parton
distribution functions} 

\author{Muhammad Ridwan\orcid{0000-0002-2949-5866}}
\email[E-mail:]{muhammad.ridwan@plymouth.ac.uk}
\affiliation{Centre for Mathematical Sciences, University of Plymouth, Plymouth PL4 8AA, United
Kingdom}

\author{Parada~T.~P.~Hutauruk\orcid{0000-0002-4225-7109}}
\email[E-mail:]{phutauruk@hiroshima-u.ac.jp} 
\affiliation{International Institute for Sustainability with Knotted Chiral Meta Matter
(WPI-SKCM$^2$), Hiroshima University, Higashi-Hiroshima, Hiroshima 739-8526, Japan}

\author{Ahmad Jafar Arifi\orcid{0000-0002-9530-8993}}
\email[E-mail:]{aj.arifi01@gmail.com} 
\affiliation{Advanced Science Research Center, Japan Atomic Energy Agency, Tokai 319-1195, Japan}
\affiliation{Research Center for Nuclear Physics (RCNP), The University of Osaka, Ibaraki 567-0047,
Japan}

\author{Kazuo Tsushima\orcid{0000-0003-4926-1829}}
\email[E-mail:]{kazuo.tsushima@gmail.com} %
\affiliation{Laboratório de Física Teórica e Computacional-LFTC, 
Programa de P\'{o}sgradua\c{c}\~{a}o em Astrof\'{i}sica e F\'{i}sica Computacional,
Universidade Cidade de S\~{a}o Paulo (UNICID), 01506-000 S\~{a}o Paulo, SP, Brazil}

\date{\today}

\begin{abstract}
In this work, we present a comprehensive investigation of the in-medium properties and internal
structure of pseudoscalar mesons within the light-front quark model (LFQM), employing both Gaussian
and power-law (PL) wave functions. We incorporate in-medium quark properties obtained from the
quark-meson coupling (QMC) model into the LFQM, thereby providing a unified description in
terms of quark degrees of freedom. We investigate the distribution amplitudes (DAs) and parton
distribution functions (PDFs) of $\pi^{\pm}$, $K^{\pm}$, $D^{\pm}$, and $B^{\pm}$ mesons in
symmetric nuclear matter. We find that the evolved in-medium pion DA and PDF are strongly suppressed
at intermediate $x$, indicating substantial medium-induced modifications to the meson structure. The
signs and magnitudes of the modified Gegenbauer coefficients $a_n'$, the parameter $\xi$, and the
Mellin moments of the DAs and PDFs in nuclear matter are found to be consistent with other
theoretical predictions. In particular, the second Gegenbauer coefficient of the pion DA, $a_2^\pi$,
increases substantially with nuclear density, from $0.096$ ($0.057$) in vacuum to $0.232$ ($0.202$)
at normal nuclear density for the Gaussian (PL) wave function, respectively. This increase
indicates a broader pion DA and an enhanced second Gegenbauer moment in the nuclear medium. We further
calculate the pion PDF ratio $R_{\rho/0}^{\pi}$ and find that its $x$ dependence exhibits a
behavior qualitatively similar to the observed European Muon Collaboration (EMC) effect,
highlighting the important role of QCD evolution in shaping the medium-modified partonic structure.
Finally, we compare the calculated free-space $\pi^{\pm}$ and $K^{\pm}$ DAs with available
free-space lattice QCD results and find reasonable agreement.
\end{abstract}

\maketitle

\section{Introduction}
Mesons, as bound states of dressed valence quarks and antiquarks, have a relatively simple internal
structure compared with nucleons. Understanding their internal structure in terms of quark and gluon
degrees of freedom provides valuable insight into quantum chromodynamics (QCD)~\cite{Berger:1979du},
the underlying theory of strongly interacting matter. The partonic structure of hadrons is encoded
in a variety of nonperturbative distributions, including generalized parton
distributions (GPDs)~\cite{Belitsky:2001ns,Diehl:2003ny}, transverse-momentum-dependent
distributions (TMDs)~\cite{Collins:1981uk,Boussarie:2023izj}, parton distribution functions
(PDFs)~\cite{Hutauruk:2016sug,Barry:2018ort,Lorce:2025aqp}, and distribution amplitudes
(DAs)~\cite{Braun:2006dg,Choi:2017uos,Arifi:2023jfe}.

In this work, we focus on DAs and PDFs of pseudoscalar mesons,
which play central roles in the description of hard exclusive and inclusive processes,
respectively. Reliable theoretical predictions for these quantities are essential for interpreting
experimental measurements in deeply virtual meson production,
Drell--Yan~\cite{Politzer:1977fi,Saclay-CERN-CollegedeFrance-EcolePoly-Orsay:1980fhh,NA3:1983ejh},
and Sullivan processes~\cite{Lu:2025bnm}. Furthermore, the endpoint behavior of meson PDFs as
$x\rightarrow1$ is of particular interest~\cite{Farrar:1975yb}, since different theoretical
approaches yield different predictions for their large-$x$ power behavior. In contrast, the
small-$x$ region, $x\rightarrow0$, is sensitive to gluon dynamics and remains comparatively poorly
understood.

Experimental data on meson DAs and PDFs remain scarce, largely because stable meson targets are not
available, although data for the pion have been obtained from Drell--Yan
measurements~\cite{E615:1989bda}. Recent advances in lattice QCD have provided valuable constraints
on meson structure; however, additional experimental data are needed to achieve a more precise
determination of pion and kaon structure. In particular, measurements of Drell--Yan and Sullivan
processes at current and future facilities can provide important complementary information. These
include the Electron-Ion Collider (EIC)~\cite{Arrington:2021biu},
AMBER/COMPASS++~\cite{Adams:2018pwt}, the 22-GeV Jefferson Lab upgrade~\cite{Accardi:2023chb},
J-PARC~\cite{Sawada:2016mao}, and the Electron-Ion Collider in China (EicC)~\cite{Anderle:2021wcy}.

Furthermore, the internal structure of hadrons, including mesons, is expected to be modified in a
nuclear medium. This phenomenon is exemplified by the European Muon Collaboration (EMC)
effect~\cite{EuropeanMuon:1981gyc,EuropeanMuon:1983wih,Arrington:2021vuu}, which provides
experimental evidence for significant modifications of nucleon structure in nuclei. Despite
extensive theoretical and experimental efforts, the origin of the EMC effect remains a long-standing
problem in hadron physics (see, e.g., Refs.~\cite{Norton:2003cb,Thomas:2018kcx,Cloet:2019mql}). In
addition to changes in internal structure, hadron properties can also be modified in the nuclear
medium, as indicated by mesonic-atom experiments~\cite{Suzuki:2002ae} and other measurements (see,
e.g., the review in Ref.~\cite{Hayano:2008vn}). Although meson DAs and PDFs have been extensively
studied in vacuum, their modifications in nuclear matter remain comparatively unexplored; see, e.g.,
Refs.~\cite{Hutauruk:2025bjd,deMelo:2016uwj,Hutauruk:2021kej,Arifi:2023jfe,Puhan:2024xdq,Hutauruk:2019ipp}.

A fully consistent framework that simultaneously describes hadron properties and internal structure
in both free space (vacuum) and nuclear matter remains unavailable. A quark-level description
is particularly important for establishing a unified connection between hadron structure and its
in-medium modification. To this end, we developed a hybrid approach~\cite{Arifi:2023jfe,Arifi:2024tix} that
combines the light-front quark model (LFQM) with the quark--meson coupling (QMC) model, thereby
incorporating relativistic quark dynamics and nuclear many-body effects within a common framework.
This approach provides a foundation for a unified description of hadron structure
in free space and in nuclear matter.

Using the LFQM-QMC approach~\cite{Arifi:2023jfe}, we investigate the DAs and PDFs of pseudoscalar
mesons, with particular emphasis on their modifications in nuclear matter. We focus primarily on the
medium-induced changes in the pion and kaon structures, while also considering heavy-light mesons to
provide a broader assessment of in-medium effects. Building on our previous
study~\cite{Arifi:2023jfe}, which examined decay constants and DAs, in the present work, we extend
the analysis by investigating the QCD evolution and Mellin moments of the DAs in nuclear matter. We
further calculate the in-medium PDFs and their Mellin moments and study their evolution under the
DGLAP equations. To assess the sensitivity to the light-front wave-function ansatz, we employ both
Gaussian and power-law (PL) wave functions, allowing us to examine model dependence, particularly in
the endpoint regions.

A key finding of this study is that the in-medium light-front wave functions (LFWFs) are enhanced
near the endpoints and suppressed in the intermediate-$x$ region as a consequence of the modified
quark properties. These changes induce a corresponding suppression of the meson DAs and PDFs at
intermediate $x$, while QCD evolution shifts the location and modifies the magnitude of this
suppression. This behavior provides a possible connection between the in-medium meson
structure and the mechanisms underlying the EMC effect.

The paper is organized as follows. Section~\ref{sec:LFQM} briefly reviews the calculation of meson
properties in the LFQM using Gaussian and PL wave functions. Section~\ref{sec:QMC} describes the
calculation of the in-medium quark properties within the QMC model, which serve as inputs for the
meson DAs and PDFs. Section~\ref{sec:da} presents the formulation of the meson DAs, including their
Gegenbauer expansion and QCD evolution, while Sec.~\ref{sec:pdf} discusses the meson PDFs, their
normalization, and QCD evolution. Numerical results for the in-medium DAs and PDFs at $\mu^2=4$ and
$27~\mathrm{GeV}^2$ are presented in Sec.~\ref{sec:num} for both wave-function ansatz. Finally,
Sec.~\ref{sec:sum} summarizes our main results and conclusions.

\section{Model Framework}
In this section, we briefly review the LFQM with Gaussian and PL wave functions for calculating meson DAs and PDFs, followed by a discussion of the in-medium quark properties obtained within the QMC model.

\subsection{Light-front quark model} \label{sec:LFQM}

Here, we begin by outlining the basic construction of the LFWF within the LFQM~\cite{Jaus:1989au,Schlumpf:1994bc,Choi:1997qh,Mukherjee:2002gb,Hwang:2010hw}. 
The meson state
$\ket{\mathcal{M}}\equiv\ket{\mathcal{M}(P,J,J_z)}$ is represented as a bound state of a constituent
quark $q$ and antiquark $\bar q$, with the total momentum $P$ and the total angular
momentum $(J,J_z)$, and is written as
\begin{eqnarray}
\ket{{\cal M}} 
&=& \int \left[ d^3\bm{p}_q \right] \left[d ^3\bm{p}_{\bar{q}} \right] 16 \pi^3 \delta^3 
\left(\bm{P}-\bm{
p}_q - \bm{p}_{\bar{q}} \right) 
\nonumber\\ && \times \mbox{} 
\sum_{\lambda_q,\lambda_{\bar{q}}} \Psi_{\lambda_q \lambda_{\bar{q}}}^{JJ_z}(x, \bm{k}_\perp)
\ket{q_{\lambda_q}(p_q) \bar{q}_{\lambda_{\bar{q}}}(p_{\bar{q}}) },
\quad\quad 
\end{eqnarray}
where $\bm{p}_i=(p_i^+,\bm{p}_{i\perp})$ and
$\left[d^3\bm{p}_i\right]\equiv d p_i^+ d^2\bm{p}_{i\perp}/(16 \pi^3)$.
The quark and antiquark momenta and helicities are denoted by
$(p_q,\lambda_q)$ and $(p_{\bar q},\lambda_{\bar q})$, respectively. The internal
light-front variables $(x,\bm{k}_\perp)$ are defined by
$x=p_q^+/P^+$ and $\bm{k}_\perp=x\bm{P}_\perp-\bm{p}_{q\perp}$.

The momentum-space LFWF of a ground-state pseudoscalar meson is given by
\begin{eqnarray}
\label{eq:5}
		\Psi^{JJ_z}_{\lambda_q\lambda_{\bar{q}} }(x, \bm{k}_{\bot}) = \Phi(x, \bm{k}_\bot)
		\  \mathcal{R}^{JJ_z}_{\lambda_q\lambda_{\bar{q}} }(x, \bm{k}_\bot),
\end{eqnarray}
where $\Phi(x,\bm{k}_\perp)$ and $\mathcal{R}^{JJ_z}_{\lambda_q\lambda_{\bar q}}(x,\bm{k}_\perp)$ denote the radial and spin-orbit wave functions, respectively. The spin-orbit wave function is obtained through the Melosh transformation~\cite{Melosh:1974cu} and can be written in covariant form as
\begin{eqnarray}
\label{eq:6}
	\mathcal{R}^{00}_{\lambda_q\lambda_{\bar{q}}} &=&  \frac{1}{\sqrt{2} \tilde{M}_0} 
	\bar{u}_{\lambda_q}^{}(p_q) \gamma_5 v_{\lambda_{\bar{q}}}(p_{\bar{q}}),
\label{M0til}	
\end{eqnarray}
with $\tilde{M}_0 \equiv \sqrt{M_0^2 - (m_q -m_{\bar{q}})^2}$,
and the invariant meson mass $M_0^2$ is defined as $M_0^2 = (\bm{k}_{\bot}^2 + m_q^2)/x  + (\bm{k}_{\bot}^2 + m_{\bar{q}}^2)/(1-x).$ 
The explicit form of the spin-orbit WFs for the pseudoscalar mesons is given by
\begin{equation}
\mathcal{R}^{00}_{\lambda_q\lambda_{\bar{q}}} (x,\bm{k}_\perp) 
=\frac{1}{\sqrt{2}\sqrt{\mathcal{A}^2 + \bm{k}_\perp^2}}
\begin{pmatrix}
-k^L 		& \mathcal{A}\\
-\mathcal{A} & -k^R\\
\end{pmatrix},
\end{equation}
with $k^{R(L)}=k_x\pm i k_y$ and $\mathcal{A}=xm_{\bar{q}} + (1-x) m_q$. It is worth noting that the
spin-orbit WFs satisfy the orthonormality condition, expressed as $\sum_{\lambda_q,\lambda_{\bar{q}}} \braket{\mathcal{R}^{JJ_z}_{\lambda_q\lambda_{\bar{q}}} }{\mathcal{R}^{J^\prime J_z^\prime}_{\lambda_q\lambda_{\bar{q}}}} = \delta_{JJ^\prime}\delta_{J_z J_z^\prime}$. For the meson ground state, the radial WF is given by
\begin{eqnarray}
\Phi_{1S}(x,\bm{k}_\perp)= \displaystyle
\sqrt{\frac{\partial k_z}{\partial x}}\,  \phi_{1S}(k),
\end{eqnarray}
where we employ two kinds of phenomenological WFs
\begin{eqnarray}
\label{Eq:WF}
\phi_{1S}(k)= 
\begin{cases}
\dfrac{4\pi^{3/4}}{\beta^{3/2}}
\exp\!\left(-\dfrac{k^2}{2\beta^2}\right) & \text{Gaussian}, \\[10pt]
\sqrt{\dfrac{128 \pi}{\beta^3}}
\left(1+\dfrac{k^2}{\beta^2}\right)^{-2}  & \text{Power-law},
\end{cases}
\end{eqnarray}
with $k=|\bm{k}|$ and $k^2 = \bm{k}_\perp^2 + k_z^2$ and the scale parameter $\beta$.
The Jacobian factor originated from the $k_z\to x$ variable transformation and is crucial to recover the rotational invariance is represented by
\begin{equation}
\frac{\partial k_z}{\partial x} = \frac{M_0}{4x(1-x)} \left[ 1 - \frac{ (m_q^2 - m_{\bar{q}}^2)^2}{M_0^4} 
\right],
\end{equation}
where $k_z(x) = \left( x - (1/2) \right) M_0 + (m^2_{\bar{q}} -m^2_q)/2M_0$.
The LFWF is subsequently normalized by
\begin{equation}
    \int \frac{\dd x \, \dd ^2 \bm{k}_\bot}{2(2\pi)^3}  \abs{ \Psi (x, \bm{k}_\bot) }^2 =1.
\end{equation}
It is worth noting that Gaussian and power-law wave functions are widely used in the LFQM~\cite{Schlumpf:1994bc,Choi:1997qh,Hwang:2010hw,Mukherjee:2002gb}. In particular, the power-law behavior of the wave function is expected to emerge in the high-momentum region as a consequence of the one-gluon-exchange interaction, where the Hamiltonian is more accurately diagonalized~\cite{Arifi:2024mff}. By contrast, the Gaussian wave function exhibits an exponential suppression at large momentum.

Furthermore, the meson states considered here are constructed in a noninteracting quark-antiquark basis following the Bakamjian--Thomas prescription~\cite{Bakamjian:1953kh}, with the interaction encoded in the radial wave function $\Phi(x,k_\perp)$. For consistency, the physical meson mass is replaced by the invariant mass, $M\to M_0$, in the relevant Lorentz factors~\cite{Choi:2013mda}. This prescription has been successfully applied to a wide range of LFQM observables; see, e.g., Refs.~\cite{Choi:2013mda,Arifi:2023uqc,Arifi:2022qnd,Choi:2024ptc,Ridwan:2024ngc}.

\subsection{Quark-meson coupling model} 
\label{sec:QMC}
In this section, we briefly review the description of nuclear matter within the QMC model, originally proposed by Guichon~\cite{Guichon:1987jp}. The QMC model has been applied extensively to nuclear matter, (hyper)nuclei, and in-medium hadron properties; see, e.g., Refs.~\cite{Saito:2005rv,Guichon:2018uew}. Here, we adopt the QMC framework used in our previous work~\cite{Arifi:2023jfe}, in which hadrons are described in terms of constituent quarks.

The standard QMC model~\cite{Guichon:1987jp,Guichon:1995ue,Saito:1996sf,Saito:2005rv,Guichon:2018uew} employs the relativistic mean-field approximation and incorporates the internal structure of the nucleon by deriving the nucleon-meson couplings from the underlying quark-meson interactions. These couplings are obtained by solving the Dirac equations for quarks and antiquarks in the presence of the mean meson fields in nuclear matter,
\begin{eqnarray}
\left[i\slashed{\partial} - (m_q - V_{q\sigma}) \mp \gamma^0 V_{q\omega}\right]
\begin{pmatrix}
\psi_q(z) \\
\psi_{\bar{q}}(z) \\
\end{pmatrix} = 0, \\
\left[i\slashed{\partial} - m_Q \right]
\begin{pmatrix}
\psi_Q(z) \\
\psi_{\bar{Q}}(z) \\
\end{pmatrix} = 0,
\end{eqnarray}
where $q$ refers to light quarks
($u$ or $d$, i.e., SU(2) flavor symmetry is assumed in this work),
and $Q$ represents heavier quarks ($s, c, b$).
Here, the MIT bag model is used to simulate quark confinement, and mean-field meson potentials are
given by $V_{q\sigma} = g_{q\sigma} \sigma,$ and $ V_{q\omega} = g_{q\omega} \omega.$ The $\rho$
meson mean field is neglected since we focus on symmetric nuclear matter.

By solving the equation above, we find the static solution for the ground-state quark or antiquark
\begin{eqnarray}
\psi(r) = \psi(\vec{r}) \exp\left(-i\varepsilon^* t/R^*\right),
\end{eqnarray}
where the eigenenergies in units of $1/R^*$ are
\begin{eqnarray}
\begin{pmatrix}
 \varepsilon^*_q \\
\varepsilon^*_{\bar{q}} \\
\end{pmatrix} &=& \Omega_q^* \pm R^* V_\omega^q,
\end{eqnarray}
with $R^*$ being the in-medium bag radius. 
Here  $\Omega_q^*
=[(x_q^*)^2 + (R^* m_q^*)^2]^{1/2}$, with $x_{q}^*$ are the lowest mode bag eigenfrequencies.
We note that the mean fields couple only to the light quarks and antiquarks. 
For heavier quarks ($Q = s, c, b$), their masses remain the same in the nuclear medium as in free space ($m_Q^* = m_Q$), since the $\sigma$ field does not couple to these heavier quarks. 

The in-medium nucleon mass and radius are computed from
\begin{eqnarray}
m_N^*(\sigma) = \frac{3\Omega^*_q - Z_N}{R^*} + \frac{4\pi R^{*3}}{3} B.
\label{hmass}
\end{eqnarray}
and the stability condition ${\dd m_N^*}/{\dd R^*} = 0.$ 
The $Z_N$ constitutes the center-of-mass and gluon correction, and $B$ is the bag constant (assumed
independent of density) to stabilize the confinement, both of which are determined simultaneously to
reproduce the experimental data of nucleon mass and radius.
 
After fixing the quark-meson couplings to reproduce the empirical properties of nuclear matter at
saturation point and solving the equations self-consistently, we obtain the scalar and vector
mean fields in nuclear matter. The scalar field modifies the effective quark mass, whereas the
vector field shifts the quark energy. The resulting in-medium constituent quark mass is given by
\begin{eqnarray}
m_q^* = m_q - V_{q}\sigma,
\end{eqnarray}
where this effective mass provides a key input for determining the in-medium quark properties
within the LFQM. Although the in-medium quark energy $\varepsilon_q^*$ also enters the formalism,
its contribution averages out over the isospin partners and is
therefore omitted.

\section{Meson Distribution Amplitudes} 
\label{sec:da}
Before discussing the QCD evolution of the distribution amplitudes, we first introduce the
pseudoscalar-meson decay constant. Its general expression is given by
\begin{equation}
    \bra{0} \bar{q}(0) \gamma^\mu \gamma_5 q(0) \ket{\mathcal{P}(P)} = i f_\mathcal{P} P^\mu.
\end{equation}
In the standard LFQM, the meson decay constant can be calculated, e.g., using a good component ($\mu=+$), as~\cite{Choi:2007yu,Arifi:2023uqc}
\begin{eqnarray}
f_\mathcal{P}^{(+)} &=& \sqrt{3} \int_0^1 \mathrm{d}x \int \frac{\mathrm{d}^2 \bm{k}_\bot}{16\pi^3}\
\frac{\Phi(x,\bm{k}_\perp) }{iP^+} \nonumber\\
&&\times \sum_{\lambda_1, \lambda_2} \mathcal{R}_{\lambda_1 \lambda_2}^{00} 
\left[\frac{\bar{v}_{\lambda_2}(p_2)}{\sqrt{1-x}} \gamma^+\gamma^5
\frac{u_{\lambda_1}(p_1)}{\sqrt{x}}\right].
\end{eqnarray}
The explicit form of $f_\mathcal{P}$ in the LFQM is given by
\begin{eqnarray}\label{eq:constant}
f_\mathcal{P}^{(+)} &=& 2\sqrt{6} \int \frac{d x\ d^2 \bm{k}_\bot}{2(2\pi)^3}  
	\frac{ {\Phi}(x, \bm{k}_\bot) }{\sqrt{\mathcal{A}^2 + \bm{k}_\bot^2}} ~\mathcal{A}.
\end{eqnarray}
The distribution amplitude $\phi(x,\mu^2)$ characterizes the leading Fock component of a hadron at
a transverse separation scale set by $\mu$. In the standard LFQM, the leading-twist DA is obtained
from~\cite{Choi:2007yu,Arifi:2023uqc}
\begin{equation}
\phi(x,\mu^2) = \frac{2\sqrt{6}}{f_\mathcal{P}^{(+)}} \int \frac{\dd^2 \bm{k}_\bot}{2(2\pi)^3}
\Phi(x, \bm{k}_\bot),
\end{equation}
and it is normalized according to $\int_0^1 \dd x \, \phi(x,\mu^2) = 1.$

\subsection{ERBL evolution}
At large renormalization scales, the evolution of the pseudoscalar-meson DA is governed by the
Efremov-Radyushkin-Brodsky-Lepage (ERBL) equation~\cite{Lepage:1980fj,Efremov:1980kz}. Its solution
can be expanded in Gegenbauer polynomials,
\begin{equation}
\label{eq:gegenbaur}
    \phi(x,\mu^2) =  6x(1-x)\sum_{n=1}^\infty C_n^{3/2}(2x -1) a_n(\mu^2),
\end{equation}
where $C_n^{3/2}$ denotes a Gegenbauer polynomial of order $3/2$, and the prime indicates that the
sum runs over even $n$ only. The Gegenbauer coefficients at the initial scale $\mu_0$ are obtained
by projecting the DA onto the corresponding Gegenbauer polynomials, yielding
\begin{equation}
    a_n(\mu_0^2) = \frac{2(2n+3)}{3(n+1)(n+2)} \int_0^1 \dd x \, C_n^{(3/2)}(2x-1) \phi(x,\mu_0^2),
\end{equation}
where we have used the orthogonality relation of the Gegenbauer polynomials. The evolved Gegenbauer
coefficients are given by
\begin{equation}
a'_n(\mu^2) = a_n(\mu_0^2) \left( \frac{\alpha_s(\mu^2)}{\alpha_s(\mu_0^2)} \right)^{
\gamma_n^{(0)}/2\beta_0},
\end{equation}
where the coupling constant is computed as $\alpha_s(\mu^2) = {4\pi}/[\beta_0 \ln \left( Q_\Lambda
\right)],$
with $Q_\Lambda={\mu^2}/{\Lambda^2_{\text{QCD}}}$, $\beta_0 = 11/3N_c - \frac{2}{3}N_f,$ and
\begin{eqnarray}
    \gamma_n &=& -2C_F \left( 3 + \frac{2}{(n+1)(n+2)}- 4\sum_{k=1}^{n+1} \frac{1}{k}\right),
\end{eqnarray}
with $C_F=4/3$, where $N_c$ and $N_f$ denote the numbers of colors and active quark flavors,
respectively. The value of $\Lambda_{\mathrm{QCD}}$ depends on the number of active flavors and the
choice of renormalization scheme. Once the evolved Gegenbauer coefficients are determined, the
evolved DA, $\phi'(x,\mu^2)$, can be reconstructed from Eq.~\eqref{eq:gegenbaur}, together with its
$\xi$ moments.
\begin{equation}
  \langle \xi^n \rangle = \int_0^1 \dd x \, \xi^n \, \phi'(x,\mu^2),
\end{equation}
where $\xi=2x-1$ characterizes the longitudinal momentum asymmetry between the constituent quark
and antiquark.

\section{Meson Parton Distribution Functions}
\label{sec:pdf}
A generic expression of the pseudoscalar-meson PDF is given by
\begin{eqnarray}
    f_q (x) &=& P^+ \int \frac{\dd\xi^2}{2\pi} e^{[ ixP^+ \xi^-]} \mel{\mathcal{P}}{\bar{\psi}_q (0) \gamma^+ \psi_q (\xi^-) }{\mathcal{P}}_c\nonumber \\
\end{eqnarray}
where $q$ denotes the quark flavor, $c$ labels the connected matrix element, and $x=k^+/P^+$ is the Bjorken variable, with $P^+$ and $k^+$ denoting the plus components of the pseudoscalar-meson momentum $P$ and the struck-quark momentum $k$, respectively. Equivalently, the PDF gives the probability density for finding a collinear quark carrying a fraction $x$ of the meson momentum at the scale $\mu^2$~\cite{Lan:2019img,Li:2017mlw,Lorce:2025aqp}. Within the LFQM, it is obtained by integrating the squared LFWF over the transverse momentum,
\begin{equation}
    f_q(x;\mu^2) = \int\frac{\dd^2\bm{k}_\perp}{16\pi^3}|\Phi(x,\bm{k}_{\perp})|^2.
\end{equation}

The pseudoscalar-meson PDFs satisfy the normalization condition
\begin{equation}
\label{eqbsm}
\begin{aligned}
    \int_0^1 \dd x\,& f_q(x;\mu^2) = 1, \\
    \int_0^1 \dd x\,& \left[x f_q(x;\mu^2) + (1-x)f_{\bar q}(x;\mu^2)\right] = 1.
\end{aligned}
\end{equation}
These conditions ensure the conservation of quark number and momentum in pseudoscalar mesons. The corresponding Mellin moments are defined as
\begin{eqnarray}
    \big< x^n \big> &=& \int_0^1 \dd x~ x^n f_q (x;\mu^2), 
\end{eqnarray}
where $n=0,1,2,\ldots$ labels the Mellin moments of the PDF. In particular, $n=1$ and $n=2$ correspond to the first and second moments, respectively. For $n=0$, the expression reduces to the quark-number sum rule in Eq.~\eqref{eqbsm}.

\subsection{DGLAP evolution}
The scale dependence of the nonsinglet quark distribution is governed by the
Dokshitzer--Gribov--Lipatov--Altarelli--Parisi (DGLAP) evolution
equation~\cite{Gribov:1972ri,Altarelli:1977zs,Dokshitzer:1977sg}, which takes the form
\begin{eqnarray}
    \label{dglap1}
\frac{\partial q_{\mathrm{NS}} (x,\mu^2)}{\partial \ln \big( \mu^2\big)} = P_{qq} (x, \alpha_s
\big( \mu^2 \big)) \otimes q_{\mathrm{NS}}(x,\mu^2),
\end{eqnarray}
where $P_{qq}$ denotes the quark-to-quark splitting kernel. The convolution of the splitting kernel
with the nonsinglet quark distribution in Eq.~\eqref{dglap1} is defined as
\begin{eqnarray}
    \label{dglap2}
P_{qq} (x, \alpha_s (\mu^2)) \otimes q_{\mathrm{NS}} (x,\mu^2) = \int _{x}^1 \frac{\dd z}{x} P
\Big(\frac{x}{z} \Big) q_{\mathrm{NS}} (z,\mu^2). \nonumber \\
\end{eqnarray}

For the singlet quark distribution, the evolution equation takes the form $q_{\mathrm{S}} (x) =
\sum_i q_i^+ = \sum_i q_i (x) + \bar{q} (x) $ with $i$ being the quark flavor. In the QCD evolution,
the singlet quark distributions can be defined by
\begin{eqnarray}
    \label{dglap3}
    \frac{\partial}{\partial \ln (\mu^2) } \Bigg[ \begin{matrix}
        q_S (x, \mu^2)  \\ g(x,\mu^2) 
    \end{matrix} \Bigg] = \Bigg[ \begin{matrix}
        P_{qq} & P_{qg} 
\\ P_{gq} & P_{gg}
\end{matrix}\Bigg] \otimes \Bigg[ \begin{matrix}
    q_{s} (x, \mu^2) \\ g(x,\mu^2) 
\end{matrix}\Bigg].
\end{eqnarray}

By solving Eq.~\eqref{dglap3} numerically, we obtain the gluon distribution in the meson. The
splitting kernels entering the QCD evolution are perturbatively expanded in powers of
$\alpha_s(\mu^2)$ as
\begin{eqnarray}
    \label{dglap4}
P (z,\mu^2) = \Bigg[ \frac{\alpha_s}{2\pi}\Bigg] P^{(0)} (z) + \Bigg[
\frac{\alpha_s}{2\pi}\Bigg]^2 P^{(1)} (z) + \cdot \cdot \cdot,
\end{eqnarray}
where the first and second terms in Eq.~\eqref{dglap4} correspond to the leading-order and
next-to-leading-order (NLO) contributions, respectively. At NLO, the strong coupling $\alpha_s$ is
given by
\begin{eqnarray}
    \label{eq:NJL-gluon19}
    \alpha_{s} = \frac{4\pi}{\beta_0\ln(Q_\Lambda)} \Bigg[ 1 - \frac{\beta_1}{\beta_0} \frac{\ln
[\ln (Q_\Lambda)]}{\ln (Q_\Lambda) }\Bigg] + \mathcal{O} \Bigg( \frac{1}{\ln^2 (Q_\Lambda)}\Bigg),
\end{eqnarray}
where $\beta_1 = 34/3 N_c^2 -10/3 N_c N_f -2 C_F N_f$. The valence-quark PDFs are then evolved
using the NLO DGLAP equations described above~\cite{Miyama:1995bd}, with the resulting
distributions discussed in Sec.~\ref{sec:num}.

\section{Numerical Results and Discussions} 
\label{sec:num}
In this section, we present numerical results for the evolved DAs and PDFs of the $\pi^\pm$,
$K^\pm$, $D^\pm$, and $B^\pm$ pseudoscalar mesons in vacuum and nuclear matter at $\mu^2=4$ and
$27~\mathrm{GeV}^2$. We also examine the corresponding Gegenbauer, $\xi$, and Mellin moments in
vacuum and at finite nuclear density. Before discussing the results, we specify the LFQM parameters
and the in-medium quark inputs obtained from the QMC model.
\begin{figure*}[t]
	\centering
	\includegraphics[width=2\columnwidth]{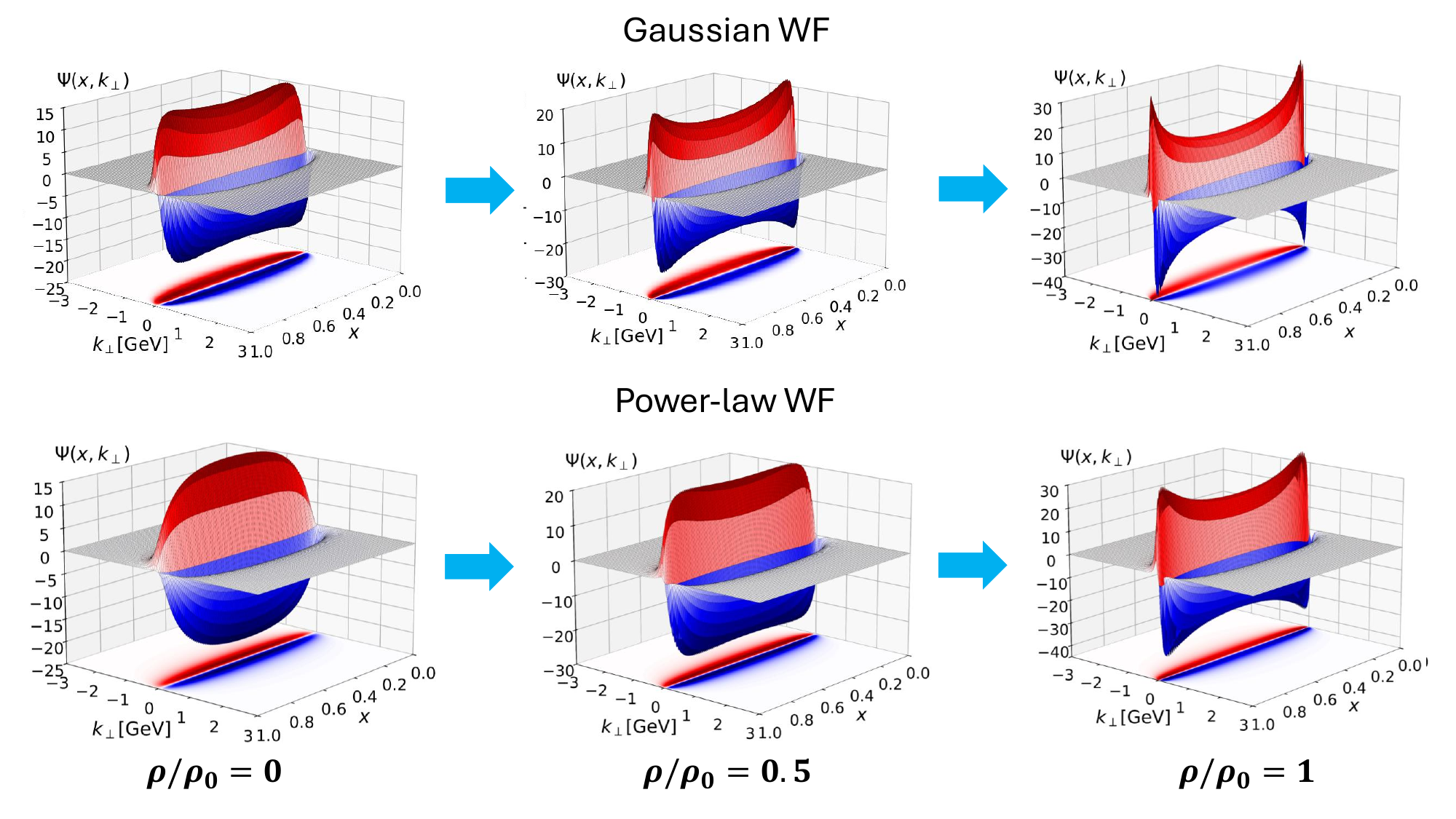}
\caption{\label{fig:LFWF_pion} LFWFs of the $\pi^\pm$ meson for the Gaussian (upper panels) and
PL (lower panels) wave functions at different nuclear densities $\rho/\rho_0$. The three-dimensional
plots show the higher-helicity components, $\Psi_{\uparrow\uparrow}=\Psi_{\downarrow\downarrow}^*$,
while the corresponding two-dimensional distributions are shown in the lower portions of each panel.
The wave function $\Psi(x,k_\perp)$ is given in units of $\mathrm{GeV}^{-1}$.}
\end{figure*}

\subsection{Model parameter}

\begin{table}[b]
    \begin{ruledtabular}
	\renewcommand{\arraystretch}{1.1}
    \caption{Constituent quark masses in vacuum and nuclear matter (in GeV), with only the
    light-quark mass $m_q$ modified in the medium ($\rho_0 = 0.15$ fm$^{-3}$)}.
    \label{tab:massparameter}
    \begin{tabular}{ccccc}
    $\rho/\rho_0$ & $m_q$ & $m_s$ &	$m_c$ & $m_b$ \\ 
    \hline
	0   & 0.22 &    &   &  \\
    0.5 & 0.14 & 0.34 & 1.8 & 5.2 \\
    1.0 & 0.08 &  &  &  \\
    \end{tabular} 
	\end{ruledtabular}
\end{table}

\begin{table}[b]
    \begin{ruledtabular}
	\renewcommand{\arraystretch}{1.1}
    \caption{Scale parameters $\beta$ for the Gaussian and PL wave functions in vacuum, given in GeV.}
    \label{tab:betaparameter}
    \begin{tabular}{ccc}
    $\mathbf{Mesons}$ & $\mathbf{\beta_{\text{G}}}$ & $\mathbf{\beta_{\text{PL}}}$ \\
    \midrule
    $\pi^{\pm}$ & 0.366 & 0.366  \\
    $K^{\pm}$   & 0.416 & 0.422 \\
    $D^{\pm}$   & 0.489 & 0.492 \\
    $B^{\pm}$   & 0.569 & 0.535 \\  
    \end{tabular}
	\end{ruledtabular}
\end{table}

Before discussing the DAs, PDFs, and their associated observables, we first determine the model
parameters. The $\beta$ parameters are fitted to the experimental decay constants listed in Table II
of Ref.~\cite{Arifi:2023jfe}. The in-medium constituent-quark masses are taken from
Ref.~\cite{Arifi:2023jfe} at $\rho/\rho_0=0$, $0.5$, and $1$ ($\rho_0 = 0.15$ fm$^{-3}$),
which we use as representative densities to examine the hybrid LFQM-QMC framework for pseudoscalar
mesons. The constituent-quark masses and $\beta$ parameters employed for the Gaussian and PL wave functions are listed in Tables~\ref{tab:massparameter} and \ref{tab:betaparameter},
respectively. The fitted $\beta$ parameters differ slightly between the two wave-function ansätze,
reflecting their different momentum dependencies. For the $\pi^\pm$ meson, $\beta_{\rm G}$ and
$\beta_{\rm PL}$ are comparable, whereas a more pronounced difference is found for the $B^\pm$
meson, for which $\beta_{\rm PL}$ is smaller than $\beta_{\rm G}$. The two parameters are also
similar for the $D^\pm$ meson. Overall, both $\beta_{\rm G}$ and $\beta_{\rm PL}$ increase
monotonically with the meson mass.

The difference between $\beta_{\mathrm{G}}$ and $\beta_{\mathrm{PL}}$ originates from the distinct
momentum dependence of the corresponding wave functions in Eq.~\eqref{Eq:WF}. These different
functional forms lead to different momentum-space profiles and, consequently, to distinct values of
the scale parameter $\beta$. The magnitude of this difference also depends on the quark flavors
constituting the meson. Similar behavior has been reported in
Refs.~\cite{Schlumpf:1994bc,Choi:1997qh}, where the PL wave function is parametrized by a power $s$
in the denominator, with $s=2$.

Using these parameters, we investigate the DAs and PDFs of the $\pi^\pm$, $K^\pm$, $D^\pm$, and
$B^\pm$ pseudoscalar mesons in nuclear matter. The in-medium up- and down-quark masses obtained from
the QMC model are used as inputs at $\rho/\rho_0=0$, $0.5$, and $1$, as listed in
Table~\ref{tab:massparameter}. For the heavy quarks, we retain their vacuum properties because their
interactions with the nuclear medium are comparatively weak. The determination of the QMC model
parameters is described in detail in Ref.~\cite{Arifi:2023jfe}.

\subsection{Light-front wave function}
In this subsection, we present three-dimensional (3D) representations of the pion LFWFs in
Fig.~\ref{fig:LFWF_pion} to illustrate their evolution in nuclear matter. For clarity, we show only
the higher-helicity components; the distinction between the higher- and ordinary-helicity components
is discussed in Ref.~\cite{Ridwan:2024ngc}. As the nuclear density increases, the magnitude and
shape of the LFWFs are substantially modified. In vacuum, the distributions exhibit a single-peak
structure, which gradually develops into a two-peak structure in nuclear matter, with the peaks
becoming increasingly separated for both the Gaussian and PL wave functions. The 3D representation
makes this transition particularly apparent, whereas it is less evident in the corresponding
two-dimensional projections. The emergence of the two-peak structure near the endpoints at higher
densities is primarily associated with the reduction of the in-medium effective quark mass, as shown
in Table~\ref{tab:massparameter}. These modifications of the LFWFs are directly reflected in the
meson structure and observables in nuclear matter. In particular, because the DAs and PDFs are
determined from the LFWFs, their in-medium modifications provide a direct link between the
underlying quark dynamics and observable meson structure.

Figure~\ref{fig:LFWF_pion} shows that the pion LFWFs obtained with the Gaussian and PL wave
functions exhibit somewhat different shapes in vacuum. As the nuclear density increases, however,
their profiles become increasingly similar, particularly at $\rho/\rho_0=1$, although the Gaussian
wave function remains more sharply peaked near the endpoints than the PL form. The difference
between the two wave function ansätze is most pronounced in the limiting momentum regimes,
$k_\perp\to\infty$ and $k_\perp\to0$. The PL wave function generally has a slower falloff at large
$k_\perp$ and a stronger enhancement at small $k_\perp$ than the Gaussian form. To better understand
this behavior, we expand $\phi_{\mathrm{G}}$ and $\phi_{\mathrm{PL}}$ in the low-momentum region,
$k_\perp\to0$, corresponding to $k_\perp^2/\beta^2<1$, using Eq.~\eqref{Eq:WF}:
\begin{eqnarray}
\label{eq:lfwf_expand}
\phi_{\text{G}} &\propto&
\left.\text{exp}\left(\frac{-k^2}{2\beta^2}\right)\right|_{k\to0} = 1 - \frac{k^2}{2\beta^2} +
\frac{k^4}{8\beta^4} + \dots,\quad \\
\phi_{\text{PL}} &\propto& \left.\left(1 + \frac{k^2}{\beta^2}\right)^{-2}\right|_{k\to0} = 1 -
\frac{2k^2}{\beta^2} + \frac{3k^4}{\beta^4} + \dots.
\end{eqnarray}

\begin{figure}[t]
	\centering
	\includegraphics[width=1\columnwidth]{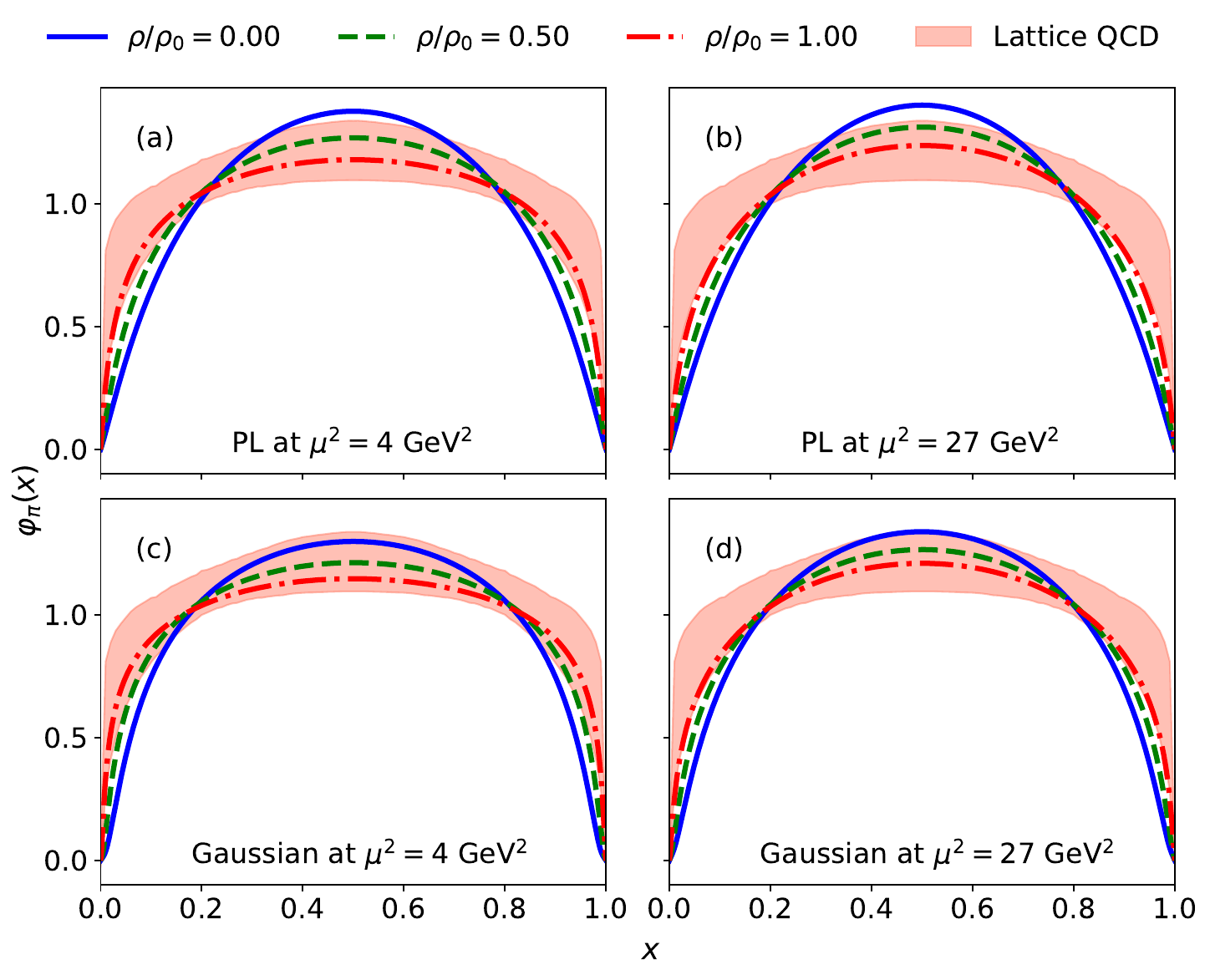}
 	\caption{\label{fig2}(a)-(d): Distribution amplitude for $\pi^{\pm}$ meson in medium
using the Gaussian and PL WFs at the scales $\mu^2 =$ 4 and 27 GeV$^2$. 
The orange shaded area represents the lattice QCD simulation result, which is taken from
Ref.~\cite{LatticeParton:2022zqc}.}
\end{figure}

Both wave functions therefore exhibit the same even-power dependence on $k/\beta$, but with
different coefficients. In particular, the larger magnitude of the leading $k^2/\beta^2$ coefficient
in the PL wave function implies a steeper initial decrease with increasing $k$. In the high-momentum
limit, $k\to\infty$, the corresponding asymptotic behaviors are given by
\begin{eqnarray}
\phi_{\text{G}} &\propto& \exp\left(-\frac{k^2}{2\beta^2}\right)\underset{k\to\infty}{\sim}
\exp\left(-\frac{k^2}{2\beta^2}\right), \\
\phi_{\text{PL}} &\propto& \left(1+\frac{k^2}{\beta^2}\right)^{-2}
\underset{k\to\infty}{\sim}
\frac{\beta^4}{k^4}.
\end{eqnarray}
Consequently, the Gaussian LFWF exhibits a somewhat faster falloff at large $\bm{k}_{\perp}$ than
the PL LFWF. These differences in the momentum dependence are expected to influence the shapes of
the $\pi^\pm$ DAs and PDFs and, more generally, those of the other pseudoscalar mesons considered
here.

\subsection{Evolved distribution amplitude}
We now present the results for the pseudoscalar-meson DAs and their QCD evolution at
$\rho/\rho_0=0$, $0.5$, and $1.0$. We compare the resulting DAs for the $\pi^\pm$ and $K^\pm$ mesons
with recent lattice QCD
calculations~\cite{LatticeParton:2022zqc,Zhang:2020gaj,Detmold:2025lyb,Zhang:2017bzy}, with
particular emphasis on the corresponding Gegenbauer coefficients. We take the initial scale to be
$\mu_0^2=1~\mathrm{GeV}^2$, chosen to provide consistency between our present work and the previous
theoretical studies, particularly for the pion and kaon DAs. The corresponding $\xi$ moments are
collected in Appendix~\ref{app:xi-moment}.

\paragraph{Pion DAs}
Figure~\ref{fig2} shows our pion DAs obtained with both LFWF ansatz in vacuum and nuclear matter
at $\mu^2=4~\mathrm{GeV}^2$ (left panels) and $27~\mathrm{GeV}^2$ (right panels). Overall, the
in-medium pion DA at $\rho/\rho_0=1$ is consistent with the lattice QCD results of
Ref.~\cite{LatticeParton:2022zqc}. As the renormalization scale increases, the pion DA becomes more
convex, approaching the asymptotic form $\phi_\pi^{\mathrm{asymp}}(x)=6x(1-x)$, as expected from QCD
evolution. The nuclear medium induces a pronounced suppression of the pion DA with increasing
density, particularly in the intermediate-$x$ region, $x\simeq0.2$--$0.8$. Conversely, the DA is
enhanced near the endpoints, $x\lesssim0.2$ and $x\gtrsim0.8$, resulting in a broader distribution
as the density increases. These features are qualitatively consistent with the unevolved results
reported in Ref.~\cite{Arifi:2023jfe}. In the present analysis, we attribute this behavior to the
modification of the LFWFs in nuclear matter, shown in Fig.~\ref{fig:LFWF_pion}, which originates
from the density dependence of the in-medium quark properties.

\begin{table}[t]
\begin{ruledtabular}
		\renewcommand{\arraystretch}{1.2}
\caption{Evolved Gegenbauer coefficients $a_n$ of the $\pi^\pm$ DA obtained with the
Gaussian and PL wave functions at $\mu^2=4~\mathrm{GeV}^2$ for different nuclear densities. Note
that the odd Gegenbauer coefficients vanish because of the symmetric quark content of the pion.
Lattice QCD results from Ref.~\cite{RQCD:2019osh} are included for comparison.}
\label{tab:a_moments_pi}
\begin{tabular}{c|ccc|c}
$a_n$ & $\rho/\rho_0 = 0.0$ & $\rho/\rho_0 = 0.5$ & $\rho/\rho_0 = 1.0$  & ~\cite{RQCD:2019osh} \\
\hline
$a_2$ &  0.096(0.057) &  0.168(0.129) &  0.232(0.202) &  $0.116^{+19}_{-20}$\\
$a_4$ &  0.002(0.001) &  0.041(0.026) &  0.093(0.071) &   - \\
\end{tabular}
	\end{ruledtabular}
\end{table}

The evolved Gegenbauer coefficients $a_n^\pi$ of the in-medium pion DA at $\mu^2=4~\mathrm{GeV}^2$
are listed in Table~\ref{tab:a_moments_pi}. Owing to the charge-conjugation symmetry of the pion,
the odd Gegenbauer coefficients vanish; therefore, only the even coefficients, such as $a_2^\pi$
and $a_4^\pi$, are considered. The vacuum value of $a_2^\pi$ is in good agreement with the
lattice QCD result~\cite{RQCD:2019osh}. For both the Gaussian and PL wave functions, the Gegenbauer
coefficients increase moderately with nuclear density, indicating that the density dependence of
the constituent-quark mass modifies the shape of the pion DA and, consequently, its $\xi$ moments.

\begin{figure}[t]
	\centering
	\includegraphics[width=1\columnwidth]{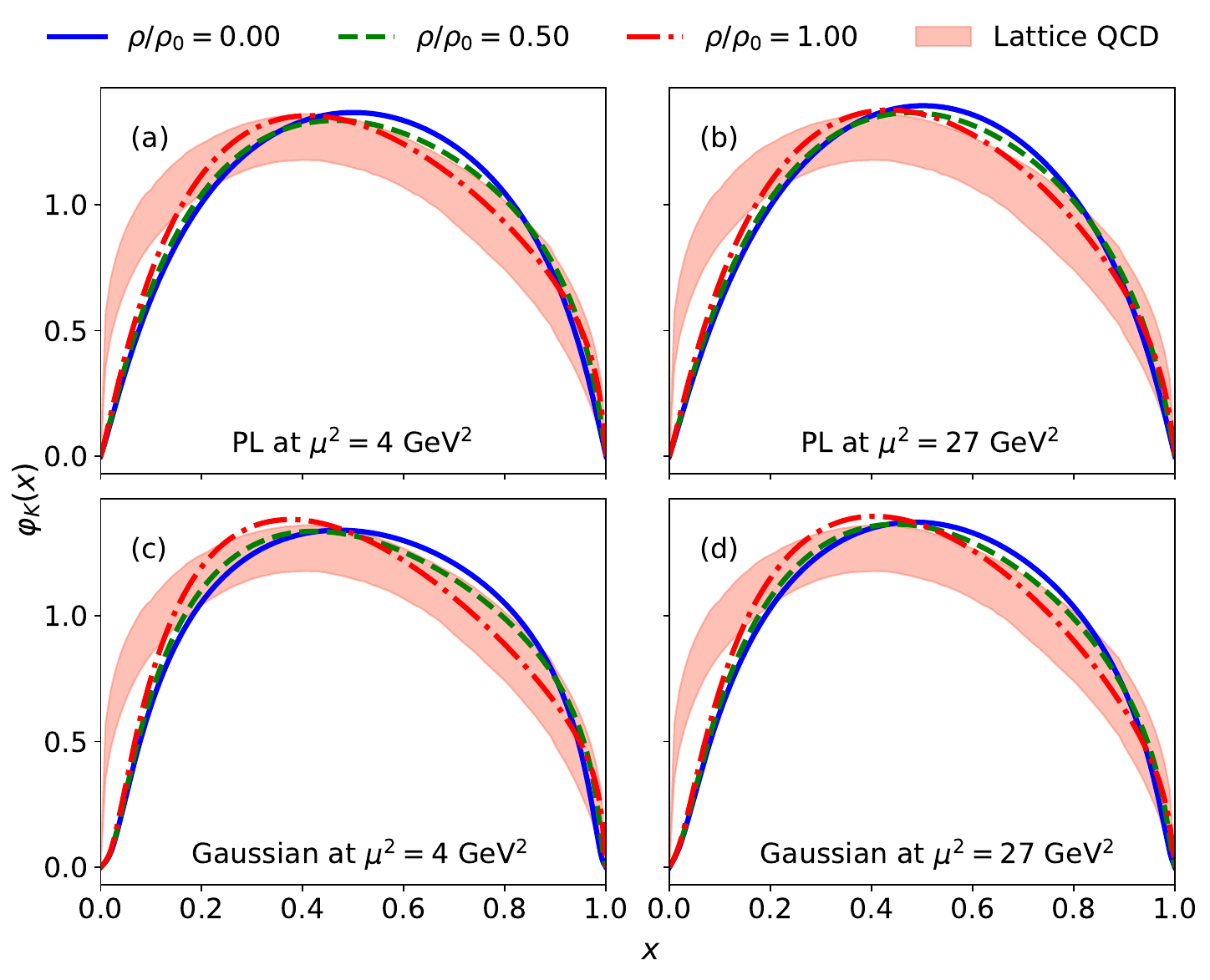}
 	\caption{\label{fig5}(a)-(d): Distribution amplitude for $K^{\pm}$ meson in medium using the
Gaussian and PL WFs at the scales $\mu^2 =$ 4 and 27 GeV$^2$. The orange shaded area represents the
lattice QCD simulation result, which is taken from Ref.~\cite{LatticeParton:2022zqc}.}
\end{figure}

\paragraph{Kaon DAs} Next, we examine the kaon DA in vacuum and nuclear matter using the Gaussian
and PL wave functions, as shown in Fig.~\ref{fig5}. Its density dependence differs from that of the
pion because of the asymmetric quark content of the kaon, particularly the presence of the heavier
strange quark, whose mass is assumed to be unaffected by the nuclear medium. In vacuum, the kaon DA
is in good agreement with the lattice QCD result~\cite{LatticeParton:2022zqc}, with the PL wave
function providing a particularly good description near the endpoints. As the nuclear density
increases, the in-medium kaon DA develops a crossing near $x\simeq0.5$: it is enhanced for
$x\lesssim0.5$ and suppressed for $x\gtrsim0.5$. This qualitative behavior persists under QCD
evolution from $\mu^2=4$ to $27~\mathrm{GeV}^2$, although the magnitude of the medium modification
is reduced at the larger scale.

\begin{figure}[t]
	\centering
	\includegraphics[width=1\columnwidth]{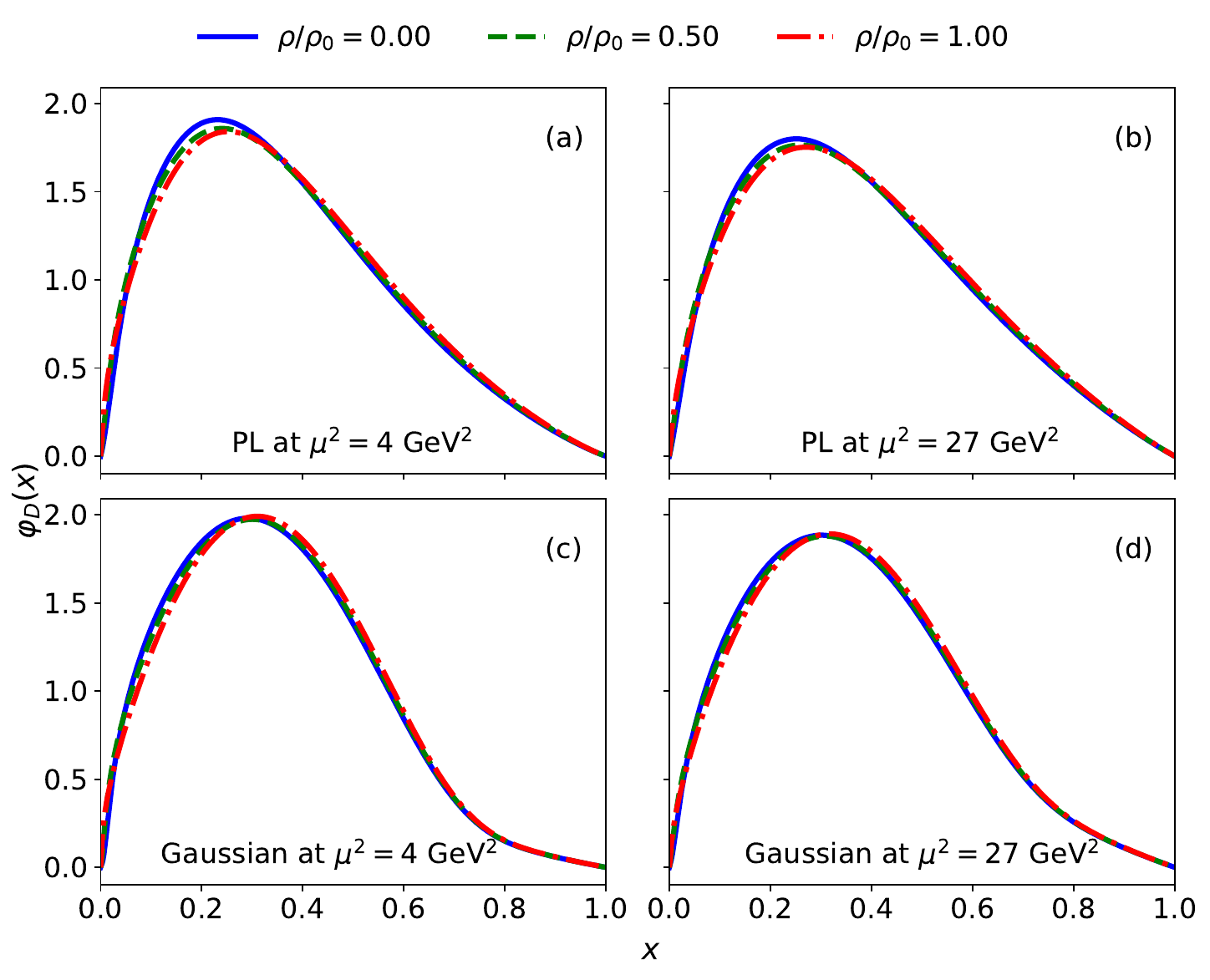}
 	\caption{\label{fig8}(a)-(d): Distribution amplitude for the $D^{\pm}$ meson in medium using the
Gaussian and PL WFs at the scales $\mu^2 =$ 4 and 27 GeV$^2$.}
\end{figure}

The evolved Gegenbauer coefficients $a_n^K$ of the kaon DA for the Gaussian and PL wave functions
at $\mu^2=4~\mathrm{GeV}^2$ are listed in Table~\ref{tab:a_moments_K}. In contrast to the pion, the
kaon has an asymmetric quark content, and its odd Gegenbauer coefficients do not vanish. In
particular, the first coefficient, $a_1^K$, provides an important measure of SU(3)-flavor-symmetry
breaking and is therefore relevant for studies of kaon DAs in QCD-based approaches. The coefficients
\begin{table}[ht]
\begin{ruledtabular}
		\renewcommand{\arraystretch}{1.2}
\caption{Evolved Gegenbauer coefficients $a_n$ of the $K^\pm$ DA for $n=1,2,3,4,$ and $5$, obtained
with the Gaussian and PL wave functions at $\mu^2=4~\mathrm{GeV}^2$ for different nuclear
densities. Lattice QCD results from Ref.~\cite{RQCD:2019osh} are included for comparison.}
\label{tab:a_moments_K}
\begin{tabular}{c|ccc|c}
$a_n$ & $\rho/\rho_0 = 0.0$ & $\rho/\rho_0 = 0.5$ & $\rho/\rho_0 = 1.0$ & ~\cite{RQCD:2019osh} \\
\hline
$a_1$ &  $-0.020(-0.038)$ &  $-0.003(-0.039)$ &   0.048($-0.003$)   &  $0.0525_{-33}^{+31}$  \\
$a_2$ &  0.072(0.043)    &   0.091(0.069)   &   0.089(0.077) & $0.106_{-16}^{+15}$ \\
$a_3$ &  $-0.028(-0.013)$ &  $-0.046(-0.030)$ &   $-0.050(-0.039)$   &  - \\
$a_4$ &  $-0.010(-0.001)$ &   0.005(0.010) &  0.010(0.020)  & - \\
$a_5$ &  $-0.010(-0.002)$ &  $-0.024(-0.009)$ &  $-0.033(-0.018)$ & - 
\end{tabular}
	\end{ruledtabular}
\end{table}

\begin{table}[ht]
\begin{ruledtabular}
		\renewcommand{\arraystretch}{1.2}
\caption{Evolved Gegenbauer coefficients $a_n$ of the $D^\pm$ DA for $n=1,2,3,4,$ and $5$, obtained
with the Gaussian and PL wave functions at $\mu^2=4~\mathrm{GeV}^2$ for different nuclear densities.
QCD sum-rule results from Ref.~\cite{Zhong:2022ugk} are included for comparison.}
\label{tab:a_moments_D}
\begin{tabular}{c|ccc|c}
$a_n$ & $\rho/\rho_0 = 0.0$ & $\rho/\rho_0 = 0.5$ & $\rho/\rho_0 = 1.0$ & ~\cite{Zhong:2022ugk}  \\
\hline
$a_1$ &  $-0.513(-0.469)$ &  $-0.507(-0.464)$ &  $-0.489(-0.446)$ & $-0.582(62)$ \\
$a_2$ &   0.088(0.141)   &   0.084(0.147)  &  0.062(0.131) & 0.148(42)  \\
$a_3$ &  $-0.002(-0.031)$ &  $-0.007(-0.048)$ & 0.006($-0.042$)   &  - \\
$a_4$ &   0.028(0.003)   &  0.040(0.022)  &  0.035(0.024) & - \\
$a_5$ &  $-0.016(0.002)$ &  $-0.028(-0.016)$ &  $-0.027(-0.021)$ & -
\end{tabular}
	\end{ruledtabular}
\end{table}
exhibit an approximately monotonic behavior with increasing $n$, consistent with the pattern found
for the lower Gegenbauer coefficients in Ref.~\cite{RQCD:2019osh}. Some higher-order coefficients
are negative; in our calculation, $a_3^K$ and $a_5^K$ remain negative as the nuclear density
increases. Similar sign patterns have been reported using the same Gegenbauer basis in
Ref.~\cite{Choi:2007yu}, where results from linear and harmonic-oscillator models were compared. The
sign of a given coefficient can be understood in terms of the unequal longitudinal-momentum
distributions carried by the strange and light quarks, which arise from their different constituent
masses~\cite{Choi:2007yu}. The negative sign of $a_1^K$ and $a_3^K$ has also been reported in the lattice QCD result in Ref.~\cite{LatticeParton:2022zqc}. The same mechanism is expected to influence the Gegenbauer coefficients of the $D^\pm$ and $B^\pm$ mesons. For the kaon, the in-medium modification of the light-quark mass leads to a slight increase in $a_1^K$, $a_2^K$, and $a_4^K$ with increasing nuclear density, whereas $a_3^K$ and $a_5^K$ decrease slightly while remaining negative.

\paragraph{$D$ meson DAs}

Next, we investigate the $D^\pm$ meson DA obtained with the Gaussian and PL wave functions at
nuclear densities $\rho/\rho_0=0$, $0.5$, and $1$, and at scales $\mu^2=4$ and $27~\mathrm{GeV}^2$.
Since the charm quark is assumed not to couple directly to the nuclear medium in the present
QMC-based framework~\cite{Arifi:2023jfe}, the medium dependence arises primarily from the
light-quark sector. The relatively large charm-quark mass further suppresses the resulting
modification of the $D^\pm$ DA. As shown in Fig.~\ref{fig8}, at $\mu^2=4~\mathrm{GeV}^2$, the DA
decreases slightly with increasing density for $x\lesssim0.3$ and increases for $x\gtrsim0.3$. This
behavior differs from that of the kaon DA and reflects the greater influence of the heavy charm
quark on the longitudinal momentum distribution of the $D^\pm$ meson. The medium dependence becomes
particularly weak as $x\to1$, and the same qualitative behavior persists at
$\mu^2=27~\mathrm{GeV}^2$.

The first five evolved Gegenbauer coefficients of the $D^\pm$ meson DA obtained with the Gaussian
and PL wave functions are listed in Table~\ref{tab:a_moments_D}. As in the kaon case, the odd
Gegenbauer coefficients are negative, whereas the even coefficients remain positive. The values of
$a_1^D$ and $a_2^D$ are in good agreement with those reported in Ref.~\cite{Zhong:2022ugk},
including their respective signs. Overall, the Gegenbauer coefficients exhibit an approximately
monotonic decrease with increasing $n$. This sign pattern persists as the nuclear density increases,
indicating that the medium modification does not alter the qualitative structure of the Gegenbauer
expansion.

\begin{figure}[t]
	\centering
	\includegraphics[width=1\columnwidth]{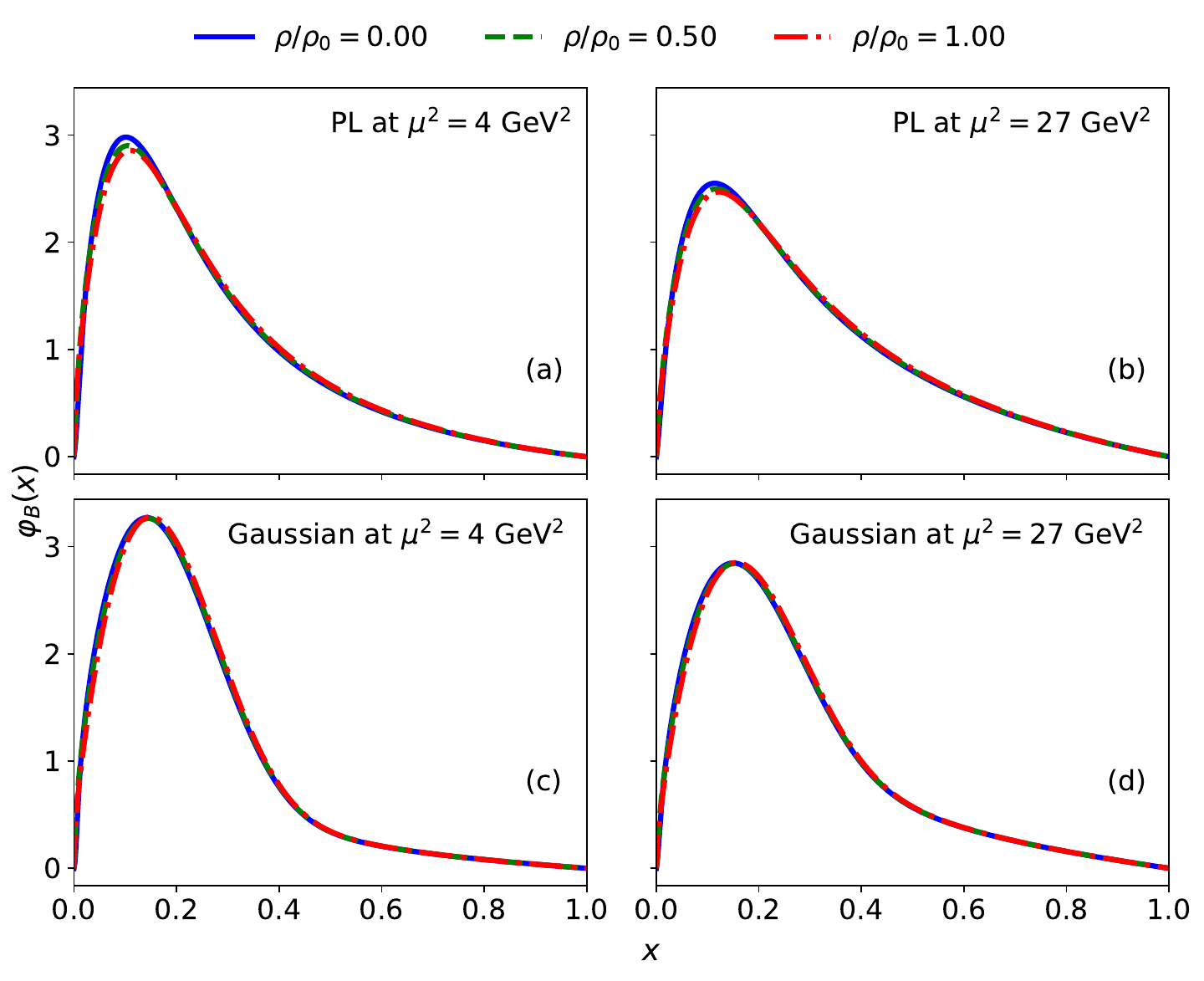}
 	\caption{\label{fig11}(a)-(d): Distribution amplitude for the $B^{\pm}$ meson in medium using
 	the Gaussian and PL WFs at the scales $\mu^2 =$ 4 and 27 GeV$^2$.}
\end{figure}

\begin{figure*}[t]
	\centering
	\includegraphics[width=1\columnwidth]{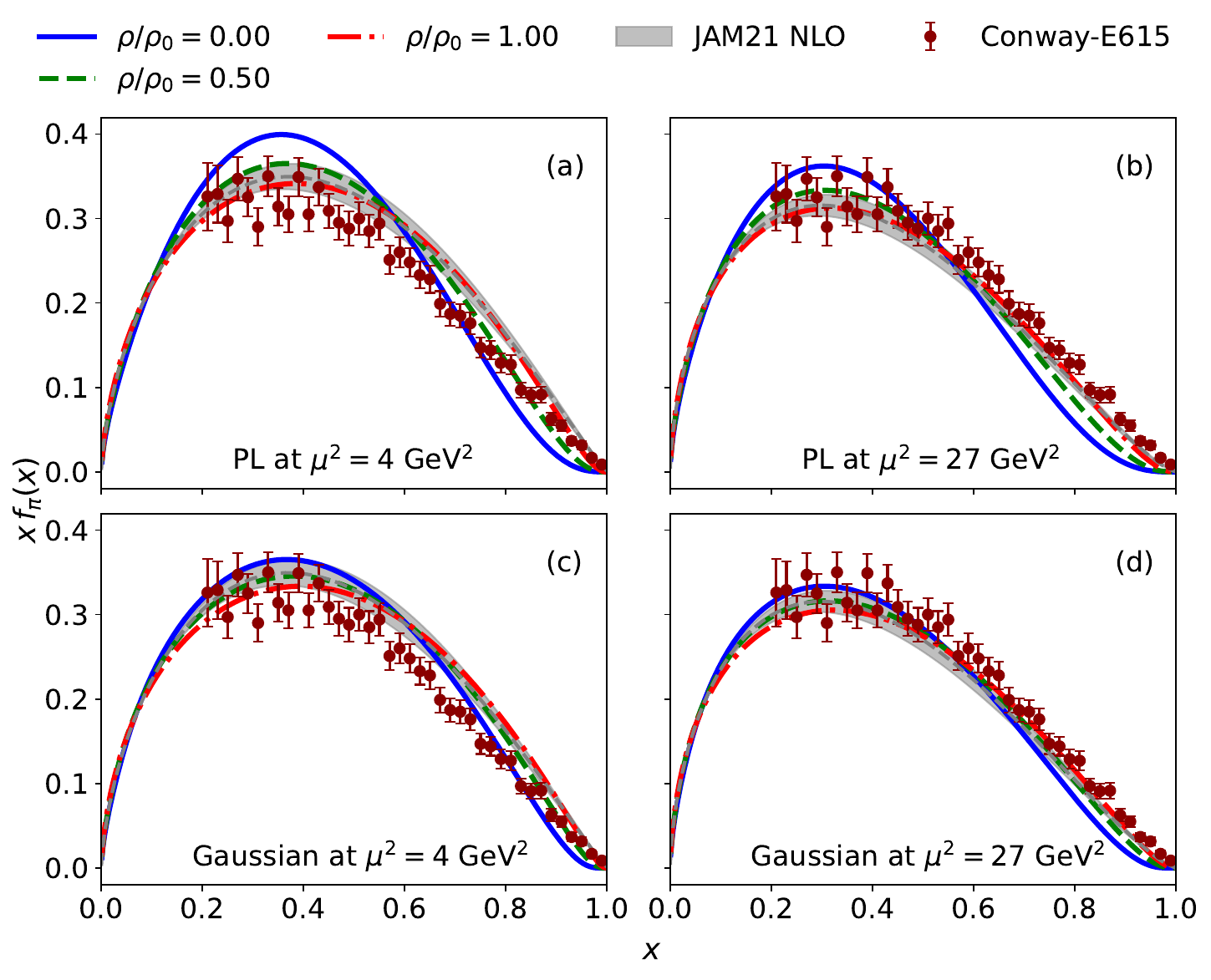}
    \includegraphics[width=1\columnwidth]{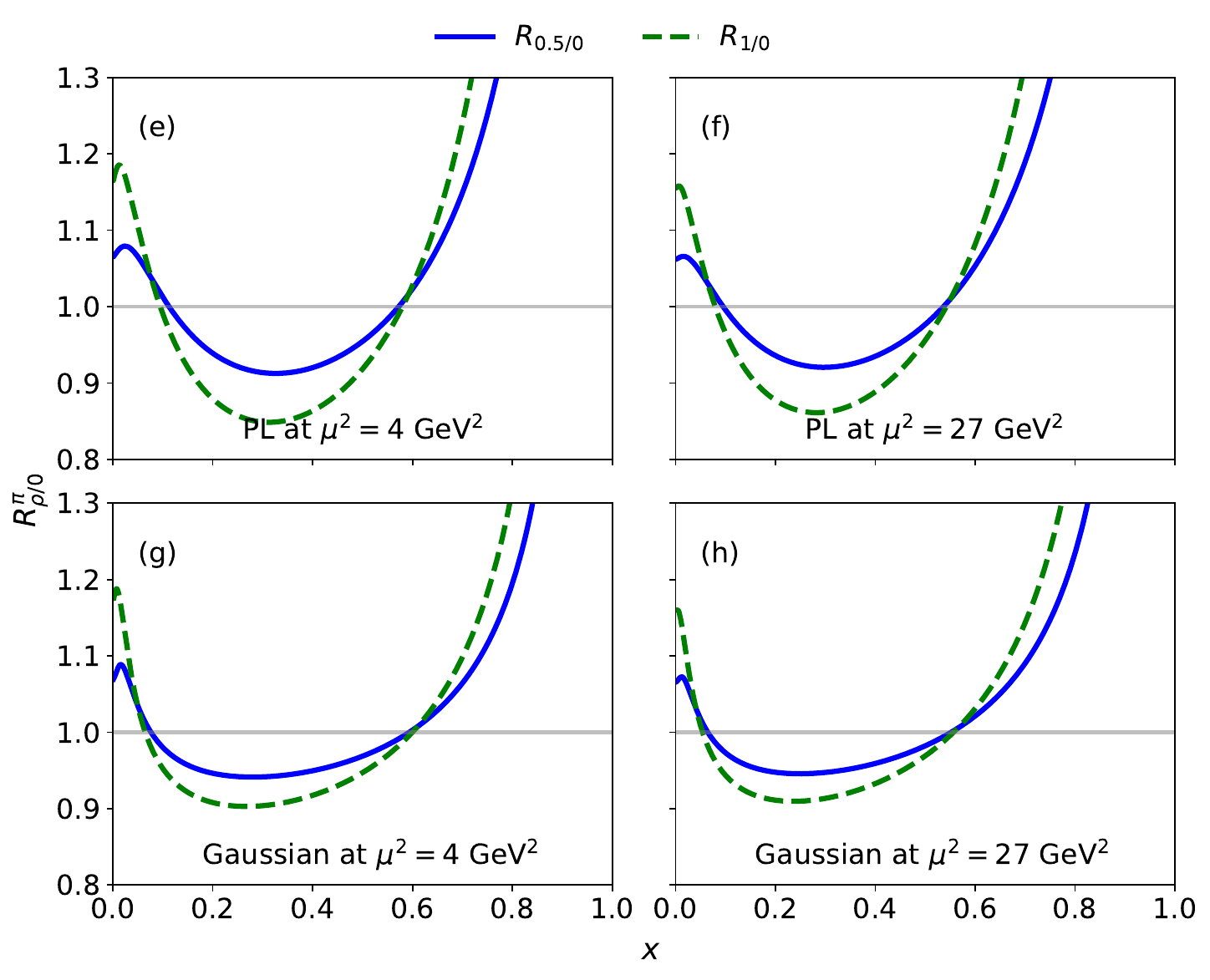}
 	\caption{\label{fig13}(a)-(d): PDFs $xf(x)$ of the $\pi^{\pm}$ meson with the PL WF at scales
$\mu^2 = 4$ (upper-left panel) and $27$ GeV$^2$ (upper-right panel) and with the Gaussian WF at the
scale $\mu^2= 4$ (lower-left panel) and $27$ GeV$^2$ (lower-right panel). For comparison, we also
present the existing experimental data (dotted points)~\cite{E615:1989bda} and JAM21 NLO analysis
(shaded line)~\cite{Barry:2025wjx}. (e)-(h): The $\pi^{\pm}$ PDF ratios $R_{\rho/0}^{\pi}$ for
respective scales and WF types which divide $\rho/\rho_0=0.5$ and $1$ (in-medium) with
$\rho/\rho_0=0$ (free space).}
\end{figure*}

\begin{figure}[htbp]
	\centering
    \includegraphics[width=1\columnwidth]{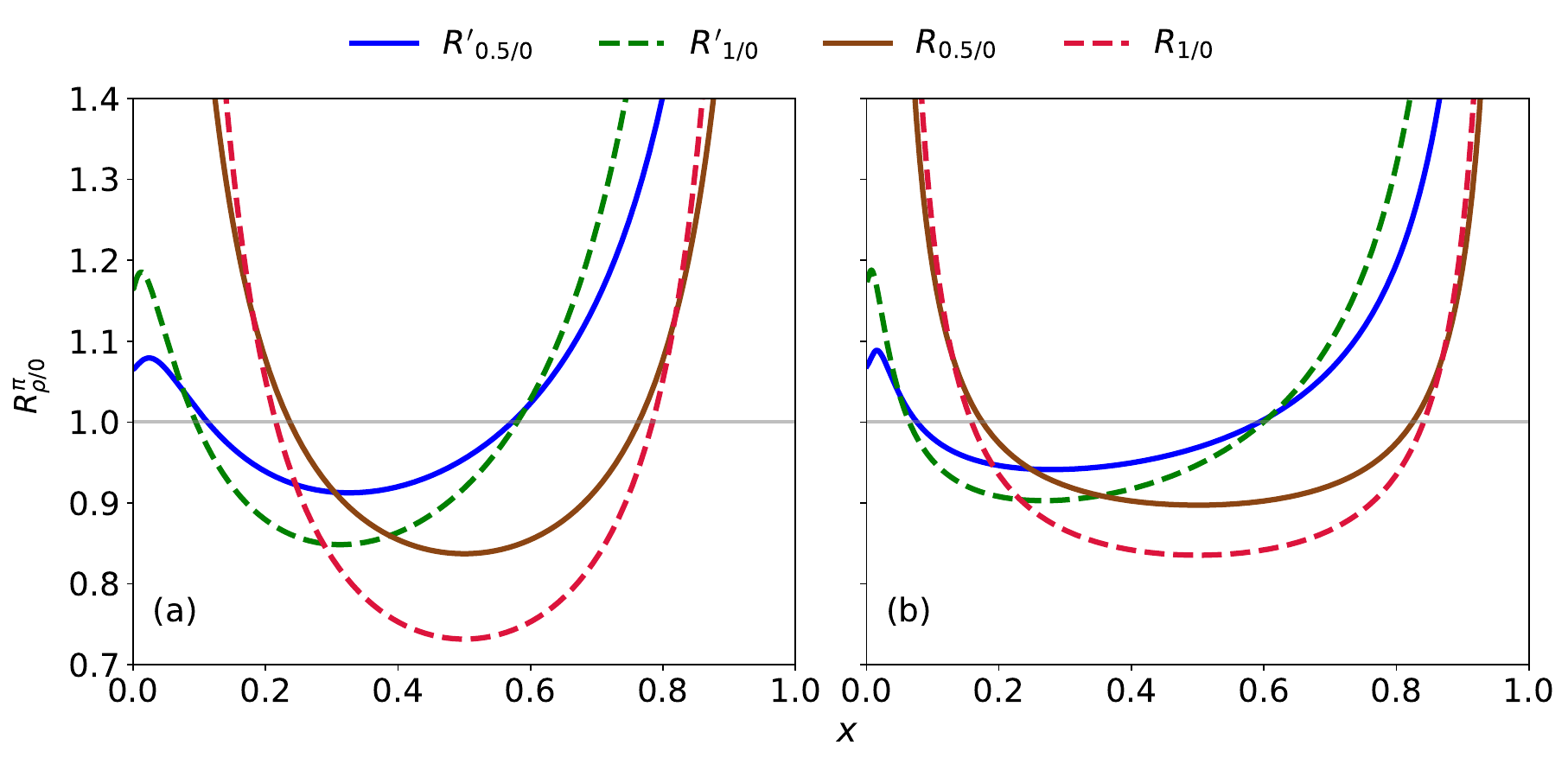}
 	\caption{\label{fig:pion_ratiocom}The comparison of the $\pi^{\pm}$ PDF ratio $R_{\rho/0}^{\pi}$
between unevolved (those with the $R$ symbols) and evolved ones (those with the $R'$ symbols) at the
scale $\mu^2=4$ GeV$^2$ in PL (a) and Gaussian (b) WFs. It clearly shows that the unevolved ratio
does have a symmetric shape, while the evolved ones have a suppression and tend to the small-$x$
region.}
\end{figure}

\paragraph{$B$-meson DAs}
We next investigate the $B^\pm$ pseudoscalar meson DA, for which the $b$ quark is
substantially heavier than the $c$ quark and is assumed to remain unaffected by the nuclear medium.
The resulting medium dependence therefore originates primarily from the light-quark sector through
modifications of its mass and momentum distribution. Owing to the large $b$-quark mass, however,
these effects have only a minor impact on the overall $B^\pm$ meson DA. As shown in
Fig.~\ref{fig11}, the medium dependence is weak at both $\mu^2=4$ and $27~\mathrm{GeV}^2$ and
becomes even less pronounced at the larger scale, consistent with the behavior observed for the
$D^\pm$ meson. The first six Gegenbauer coefficients of the $B^\pm$ meson DA obtained with the
Gaussian and PL wave functions at $\mu^2=4~\mathrm{GeV}^2$ are listed in
Table~\ref{tab:a_moments_B}. As in the $D^\pm$ case, the odd Gegenbauer coefficients are negative,
whereas the even coefficients are positive.

\begin{figure*}[t]
	\centering
	\includegraphics[width=1\columnwidth]{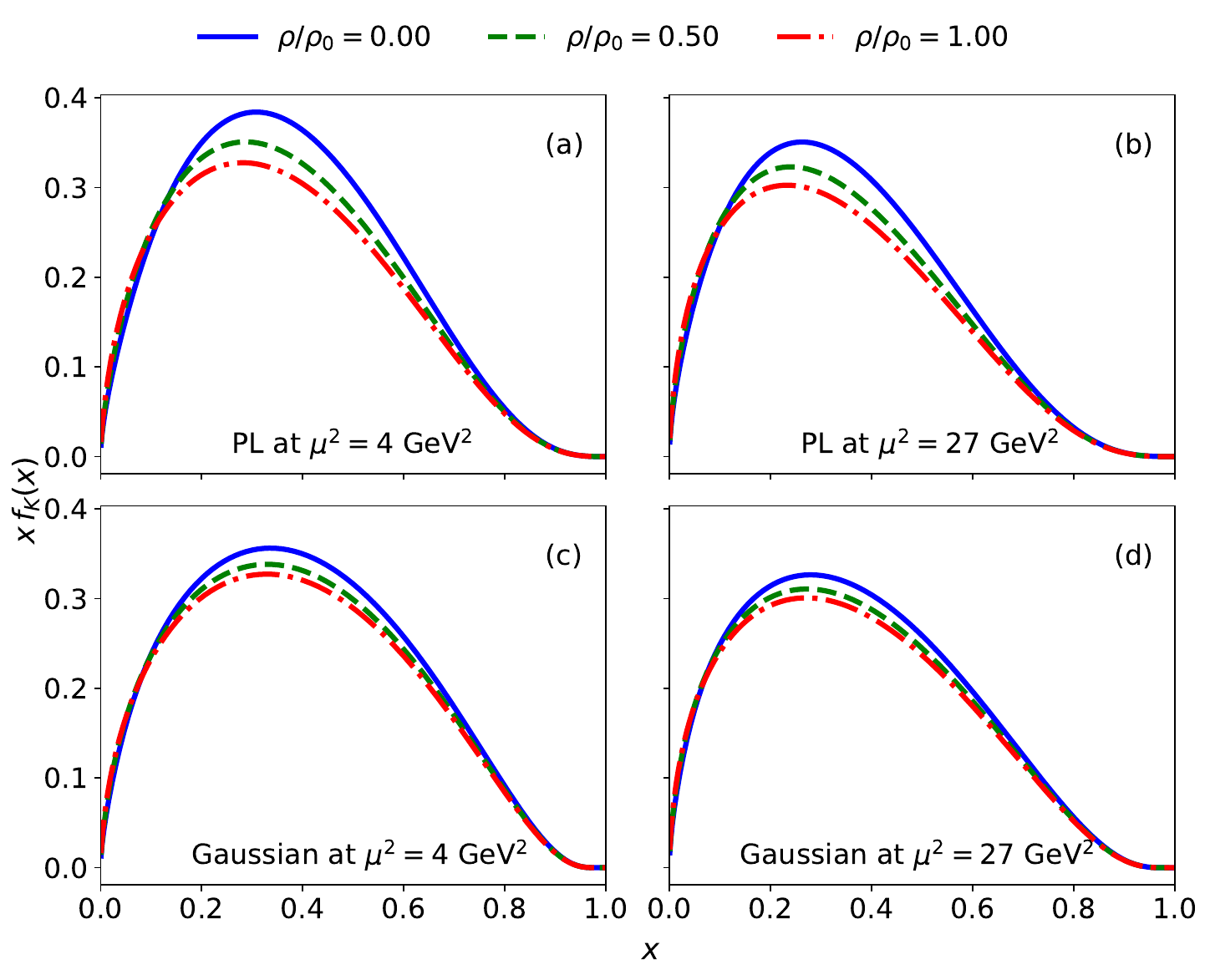}
    \includegraphics[width=1\columnwidth]{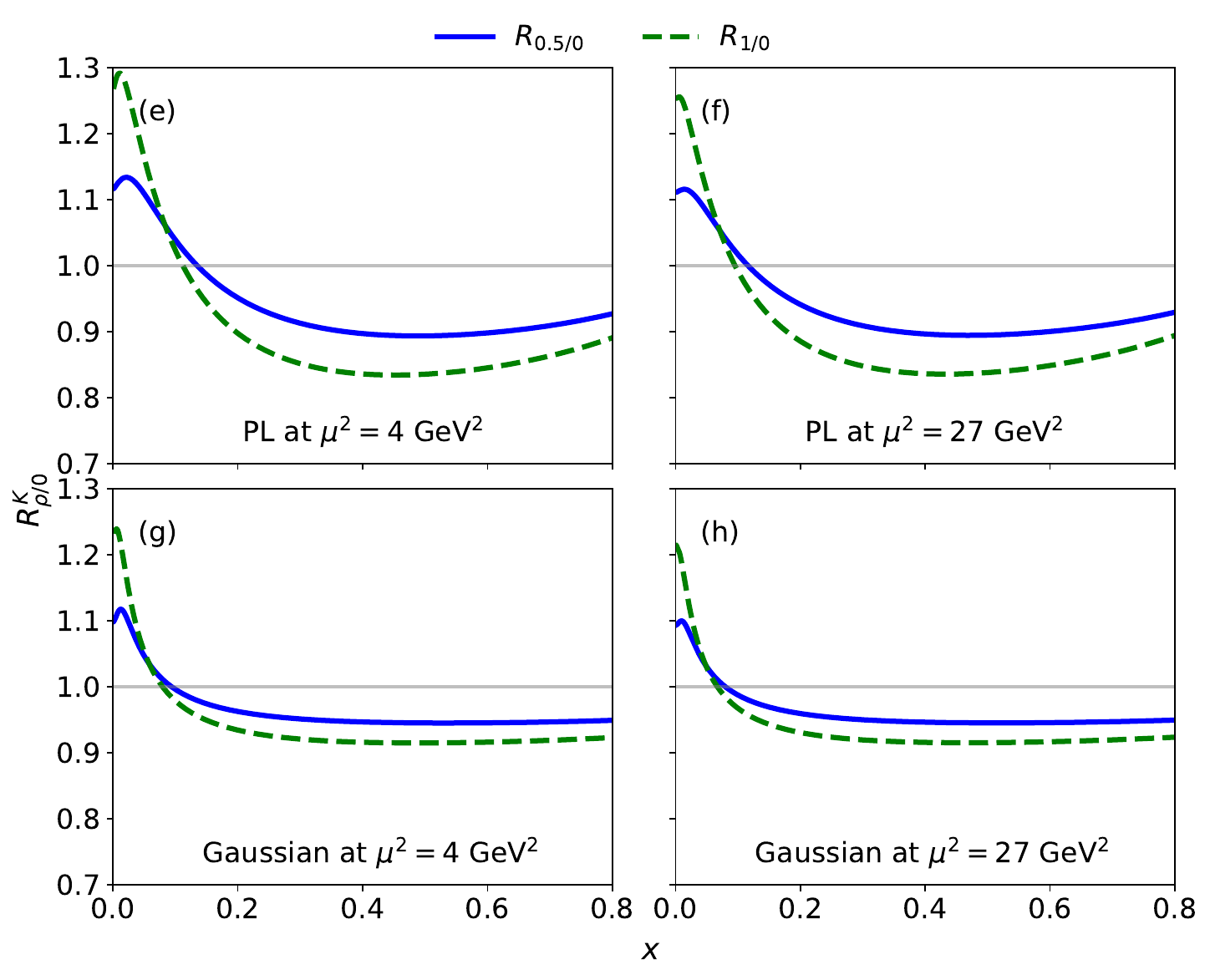}
 	\caption{\label{fig14}(a)-(d): PDFs $xf(x)$ of the $K^{\pm}$ meson with the PL WF at the scale
$\mu^2 = 4$ (upper-left panel) and $27$ GeV$^2$ (upper-right panel) and with the Gaussian WF at
scales $\mu^2 = 4$ (lower-left panel) and $27$ GeV$^2$ (lower-right panel). (e)-(h): The $K^{\pm}$
PDF ratios $R_{\rho/0}^{K}$ for respective scales and WF types which divide $\rho/\rho_0=0.5$ and
$1$ (in-medium) with $\rho/\rho_0=0$ (free space).}
\end{figure*}

\begin{table}[ht]
\begin{ruledtabular}
\renewcommand{\arraystretch}{1.2}
\caption{Evolved Gegenbauer coefficients $a_n$ of the $B^\pm$ DA for $n=1,2,3,4,$ and $5$, obtained with the Gaussian and PL wave functions at $\mu^2=4~\mathrm{GeV}^2$ for different nuclear densities.}
\label{tab:a_moments_B}
\begin{tabular}{c|ccc}
$a_n$ & $\rho/\rho_0 = 0.0$  & $\rho/\rho_0 = 0.5$  & $\rho/\rho_0 = 1.0$  \\
\hline
$a_1$ &  $-0.922(-0.816)$ &  $-0.916(-0.807)$ &  $-0.907(-0.794)$  \\
$a_2$ &   0.589(0.528)   &   0.580(0.521)   &   0.562(0.501)  \\
$a_3$ &  $-0.252(-0.293)$ &  $-0.242(-0.291)$ &  $-0.220(-0.272)$  \\
$a_4$ &   0.063(0.140)   &   0.056(0.146)   &   0.035(0.130)  \\
$a_5$ &  $-0.013(-0.058)$ &  $-0.011(-0.069)$ &   0.005($-0.058$) 
\end{tabular}
	\end{ruledtabular}
\end{table}

\subsection{Evolved parton distribution function}
\begin{figure*}[ht]
	\centering
	\includegraphics[width=1\columnwidth]{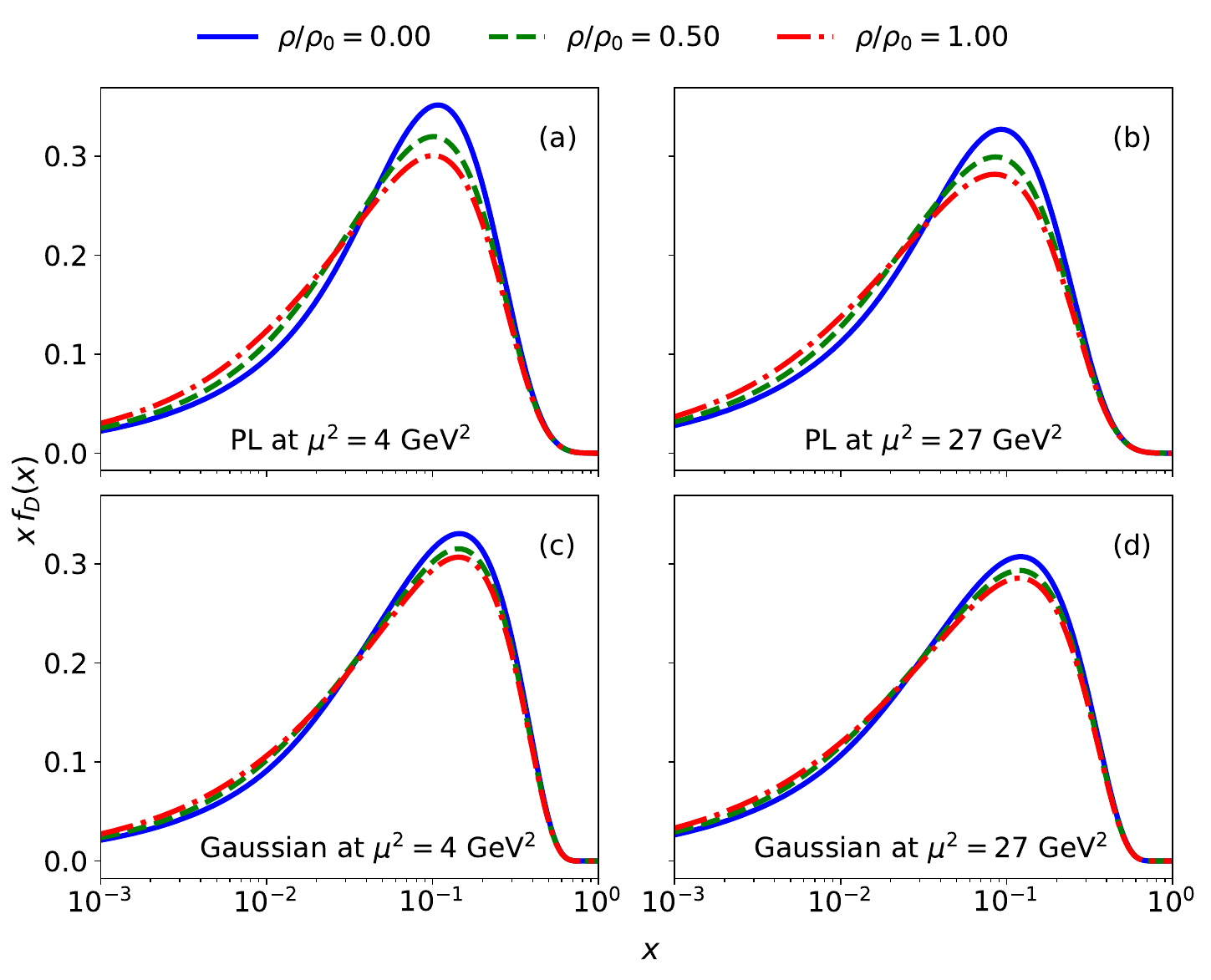}
    \includegraphics[width=1\columnwidth]{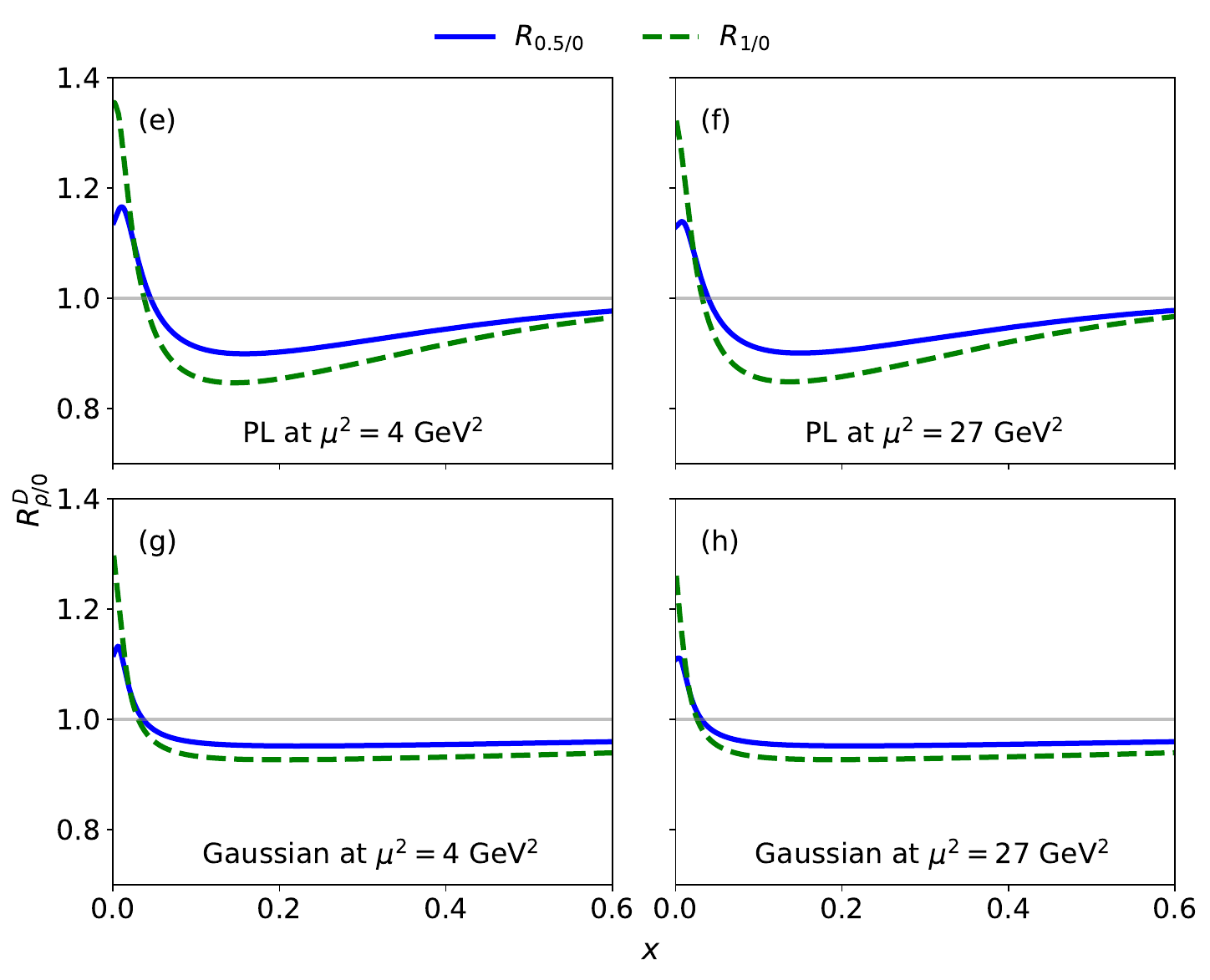}
 	\caption{\label{fig15} (a)-(d): PDFs $xf(x)$ of the $D^{\pm}$ meson with the PL WF at scales
$\mu^2 = 4$ (upper-left panel) and $27$ GeV$^2$ (upper-right panel) and with the Gaussian WF at
scales $\mu^2 = 4$ (lower-left panel) and $27$ GeV$^2$ (lower-right panel). (e)-(h): The $D^{\pm}$
PDF ratios $R_{\rho/0}^{D}$ for respective scales and WF types which divide $\rho/\rho_0=0.5$ and
$1$ (in-medium) with $\rho/\rho_0=0$ (free space).}
\end{figure*}

\begin{figure*}[ht]
	\centering
	\includegraphics[width=1\columnwidth]{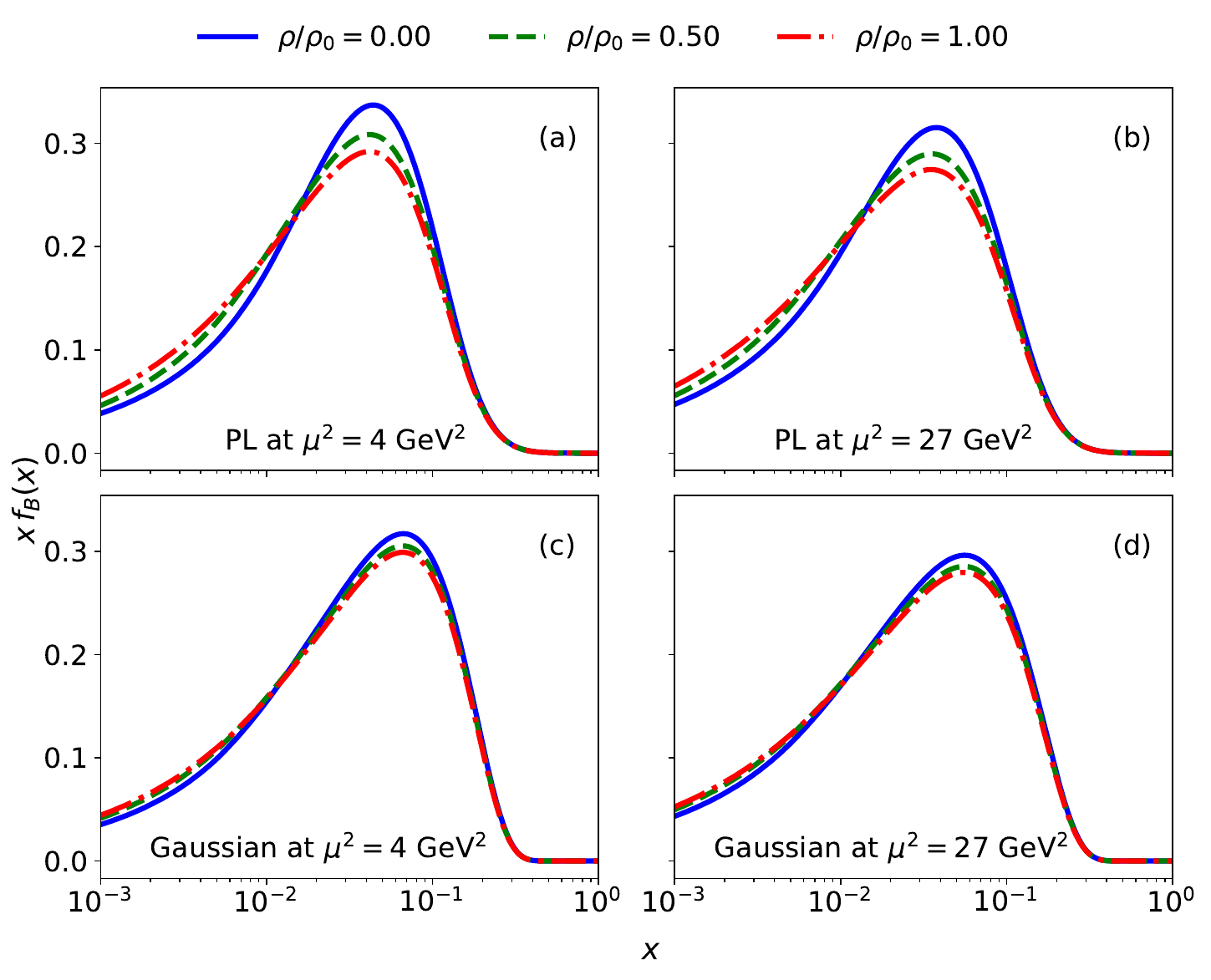}
    \includegraphics[width=1\columnwidth]{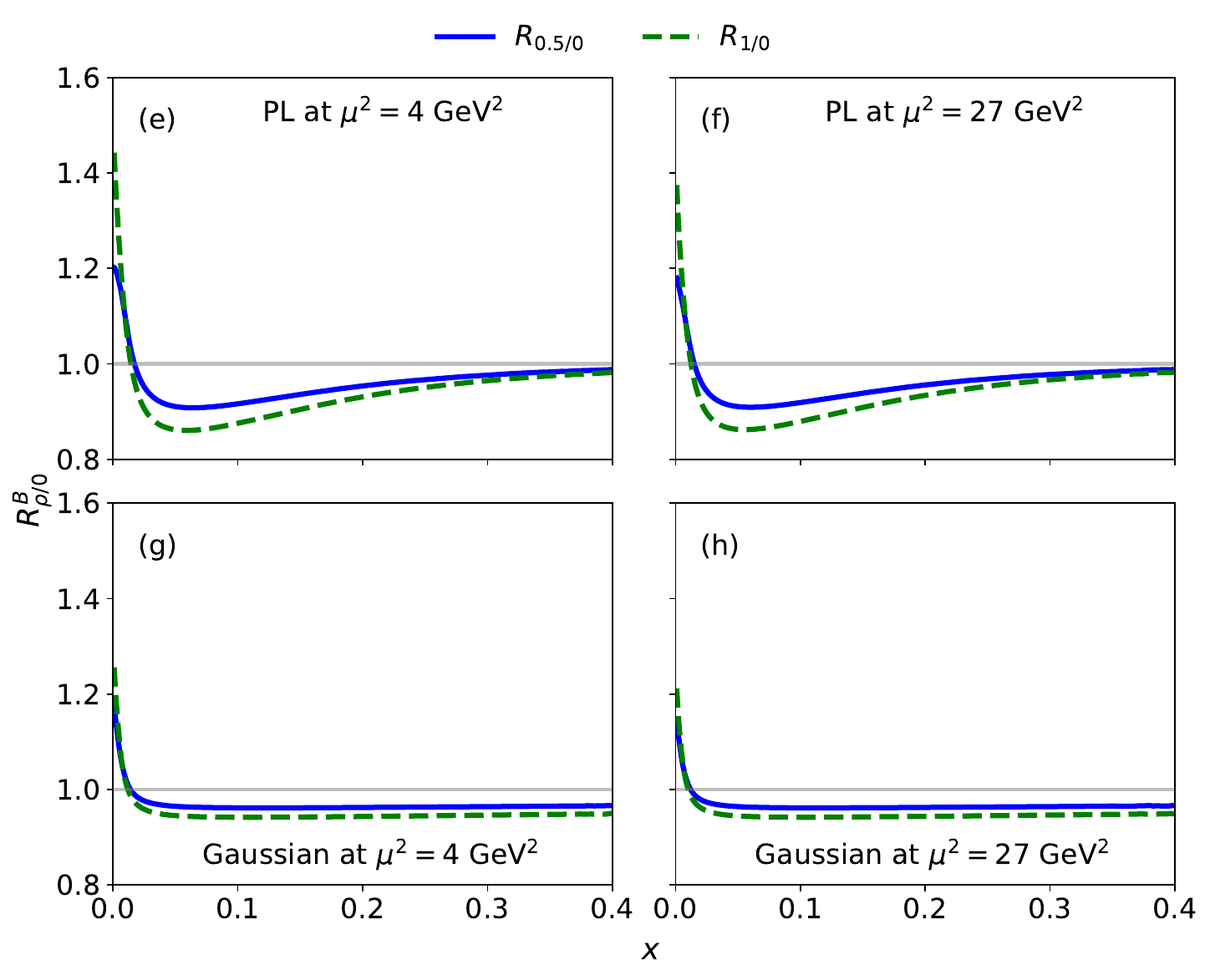}
 	\caption{\label{fig16} (a)-(d): PDFs $xf(x)$ of the $B^{\pm}$ meson with the PL WF at scales
$\mu^2 = 4$ (upper-left panel) and $27$ GeV$^2$ (upper-right panel) and with the Gaussian WF at
scales $\mu^2 = 4$ (lower-left panel) and $27$ GeV$^2$ (lower-right panel). (e)-(h): The $B^{\pm}$
PDFs ratio $R_{\rho/0}^{B}$ for respective scales and WF types which divide $\rho/\rho_0=0.5$ and
$1$ (in-medium) with $\rho/\rho_0=0$ (free space).}
\end{figure*}

Here, we present numerical results for the pion, kaon, $D^\pm$, and $B^\pm$ meson PDFs obtained
with the Gaussian and PL wave functions after NLO DGLAP evolution~\cite{Miyama:1995bd} to $\mu^2=4$
and $27~\mathrm{GeV}^2$. The results at different nuclear densities are shown in
Figs.~\ref{fig13}--\ref{fig16} and compared with available lattice QCD calculations and experimental
data.

For the PDF analysis, in contrast to the DA analysis, we choose the initial scale
$\mu_0^2=0.18~\mathrm{GeV}^2$ to obtain consistency with the pion PDF data from the Conway
experiment~\cite{E615:1989bda}, as shown in Fig.~\ref{fig13}. We also examine the ratio
$R_{\rho/0}(x)= f(x,\rho)/f(x,0)$, where $f(x,\rho)$ and $f(x,0)$ denote the in-medium and vacuum
PDFs, respectively. To further characterize the medium modification, we compare the unevolved and
evolved PDFs. This comparison is motivated by the EMC effect, which is commonly quantified through
the ratio of deep-inelastic-scattering cross sections for nuclei relative to
deuterium~\cite{Beminiwattha:2023est}. Since deuterium is only weakly bound, it provides an
approximate reference for nearly free nucleons. The analogous ratio considered here therefore offers
a useful measure of how nuclear matter modifies the internal partonic structure of pseudoscalar
mesons.

\paragraph{Pion PDFs}
Figure~\ref{fig13} shows our pion PDFs in vacuum and nuclear matter at $\mu^2=4~\mathrm{GeV}^2$
(left panels) and $27~\mathrm{GeV}^2$ (right panels), together with the experimental data of Conway
et al.~\cite{E615:1989bda} and Jefferson Lab~\cite{Barry:2025wjx}. For the latter, the data are
evolved to the corresponding scales for comparison. At both scales, the calculated pion PDFs in
vacuum and nuclear matter are consistent with the Jefferson Lab results. At
$\mu^2=27~\mathrm{GeV}^2$, the evolved PDFs also agree well with the Conway data for $x\gtrsim0.4$,
consistent with the behavior found in the NJL model~\cite{Chandra:2025pqs,Chandra:2026smf}. As the
nuclear density increases, the pion PDF is suppressed in the intermediate-$x$ region, approximately
$x\simeq0.2$--$0.6$. For $x\gtrsim0.6$, the density dependence changes, indicating a distinct
modification of the pion's large-$x$ structure in nuclear matter.

The evolved Mellin moments $\langle x^n\rangle$ of the in-medium pion PDF at
$\mu^2=4~\mathrm{GeV}^2$ are summarized in Table~\ref{tab:mellin_pion}. Overall, the moments
decrease monotonically with increasing $n$, in qualitative agreement with the lattice QCD results of
Ref.~\cite{Alexandrou:2026nsl}. The moments obtained with the Gaussian and PL wave functions remain
very similar across the considered nuclear densities, indicating only a mild sensitivity to the
wave-function ansatz. This behavior reflects the symmetric modification of the two light-quark
constituents in the pion. Accordingly, we assume identical Mellin moments for the light quark and
antiquark.

\begin{table}[ht]
\begin{ruledtabular}
		\renewcommand{\arraystretch}{1.2}
\caption{Evolved Mellin moments $\langle x^n\rangle_q$ of the $\pi^\pm$ PDF obtained with the
Gaussian (PL) wave functions at $\mu^2=4~\mathrm{GeV}^2$ for different nuclear densities.
Lattice QCD results from Ref.~\cite{Alexandrou:2026nsl} are included for comparison.}
\label{tab:mellin_pion}
\begin{tabular}{c|ccc|c}
$\langle x^n \rangle_q$ & $\rho/\rho_0 = 0.0$  & $\rho/\rho_0 = 0.5$  & $\rho/\rho_0 = 1.0$ &
\cite{Alexandrou:2026nsl} \\
\hline
$\langle x^{1} \rangle_q$ &  0.232(0.232) &  0.232(0.232) &  0.232(0.232) & 0.2194(22)  \\
$\langle x^{2} \rangle_q$ &  0.098(0.094) &  0.101(0.098) &  0.103(0.102) & 0.1021(34)  \\
$\langle x^{3} \rangle_q$ &  0.052(0.047) &  0.055(0.052) &  0.058(0.056)  & 0.079(25) \\
$\langle x^{4} \rangle_q$ &  0.031(0.027) &  0.034(0.031) &  0.037(0.035) & - \\
$\langle x^{5} \rangle_q$ &  0.020(0.017) &  0.023(0.020) &  0.026(0.024) & - 
\end{tabular}
	\end{ruledtabular}
\end{table}

Furthermore, the PDF ratios $R_{\rho/0}$ for the $\pi^\pm$ meson, defined as the ratio of the
in-medium to vacuum distributions, are shown in Fig.~\ref{fig13}(e)--\ref{fig13}(h), with the panel
arrangement corresponding to that of the PDFs. We consider the two density ratios $
R_{0.5/0}(x)=f(x,0.5)/f(x,0)$ and $R_{1/0}(x)=f(x,1)/f(x,0)$. At small $x$, the PL wave function
yields ratios closer to unity at both scales than the Gaussian wave function, indicating weaker
medium modifications. Toward larger $x$, the ratios develop a broad suppression region followed by
an enhancement. The detailed behavior depends on the wave-function ansatz, with the PL wave function
exhibiting a more pronounced deviation from unity over approximately $x=0.1$--$0.6$. Qualitatively,
this suppression-and-enhancement pattern resembles the EMC effect observed in carbon and nitrogen
nuclei~\cite{Beminiwattha:2023est}. A similar qualitative behavior has been reported
in the QMC model studies
of both unpolarized and polarized EMC effects in the nucleon~\cite{Tronchin:2018mvu}, although the
evolution scales considered there differ from those used here.

\begin{table}[ht]
\begin{ruledtabular}
		\renewcommand{\arraystretch}{1.2}
\caption{Evolved Mellin moments $\langle x^n\rangle_q$ of the $K^\pm$ PDF obtained with the Gaussian (PL) wave functions at $\mu^2=4~\mathrm{GeV}^2$ for different nuclear densities. Lattice QCD
results from Ref.~\cite{Alexandrou:2026nsl} are included for comparison.}
\label{tab:mellin_kaon}
\begin{tabular}{c|ccc|c}
$\langle x^n \rangle_q$ & $\rho/\rho_0 = 0.0$  & $\rho/\rho_0 = 0.5$  & $\rho/\rho_0 = 1.0$ &
\cite{Alexandrou:2026nsl} \\
\hline
$\langle x^{1} \rangle_q$ &  0.217(0.210) &  0.208(0.196) &  0.202(0.186) & 0.2151(12) \\
$\langle x^{2} \rangle_q$ &  0.061(0.064) &  0.059(0.061) &  0.058(0.059) & 0.0925(11) \\
$\langle x^{3} \rangle_q$ &  0.024(0.026) &  0.024(0.026) &  0.023(0.025) & 0.0557(54) \\
$\langle x^{4} \rangle_q$ &  0.012(0.013) &  0.012(0.013) &  0.012(0.013)  & - \\
$\langle x^{5} \rangle_q$ &  0.006(0.007) &  0.007(0.007) &  0.007(0.007)  & -
\end{tabular}
	\end{ruledtabular}
\end{table}

\begin{table}[ht]
    \begin{ruledtabular}
    \renewcommand{\arraystretch}{1.2}
\caption{Mellin-moment ratios $R_n=\langle x^n\rangle_q/\langle x^n\rangle_s$ of the $K^\pm$ meson
obtained with the Gaussian (PL) wave function.}
\label{tab:mellin_ratios_kaon}
\begin{tabular}{cccc}
    $n$ & $R_n$ ($\rho = 0.0$) & $R_n$ ($\rho = 0.5$) & $R_n$ ($\rho = 1.0$) \\
    \hline
    1 & 1.002 (1.001) & 1.002 (1.001) & 1.002 (1.001) \\
    2 & 0.708 (0.819) & 0.723 (0.858) & 0.728 (0.872) \\
    3 & 0.557 (0.707) & 0.578 (0.769) & 0.586 (0.792) \\
    4 & 0.469 (0.632) & 0.496 (0.711) & 0.506 (0.743) \\
    5 & 0.413 (0.579) & 0.445 (0.672) & 0.457 (0.713) \\
\end{tabular}
    \end{ruledtabular}
\end{table}

To further assess the impact of QCD evolution on the pion PDF ratio, we compare the unevolved and
evolved results for both wave-function ansätze in Fig.~\ref{fig:pion_ratiocom}. We focus on the pion
because its PDF ratio shows the closest qualitative resemblance to the EMC effect observed in the
nucleon. For convenience, we denote the unevolved ratio by $R$ and the evolved ratio at
$\mu^2=4~\mathrm{GeV}^2$ by $R'$. The unevolved ratios are approximately symmetric, whereas QCD
evolution induces an asymmetric shape and shifts the suppression region toward smaller $x$, with
little dependence on the choice of wave function. In both cases, the ratios exhibit sizable
deviations from unity, indicating substantial medium modifications and remaining qualitatively
consistent with previous studies.

\paragraph{Kaon PDFs}
Next, we examine the kaon PDF in vacuum and nuclear matter at $\mu^2=4~\mathrm{GeV}^2$ (left
panels) and $27~\mathrm{GeV}^2$ (right panels), as shown in Fig.~\ref{fig14}. The overall density
dependence is similar to that of the pion PDF, although the kaon distribution vanishes at
approximately $x\simeq0.9$. With increasing nuclear density, the kaon PDF is suppressed in the
intermediate-$x$ region, $x\simeq0.2$--$0.6$, while a different density dependence emerges for
$x\gtrsim0.6$. The PL wave function produces a broader separation between the vacuum and in-medium
distributions than the Gaussian ansatz, consistent with the behavior observed for the pion PDF.

Furthermore, the PDF ratios $R_{\rho/0}$ for the $K^\pm$ meson are shown in
Fig.~\ref{fig14}(e)--(h). Unlike the pion case, the kaon PDFs approach zero near $x\simeq0.9$,
causing the ratio to become ill-defined as both the in-medium and vacuum distributions vanish. We
therefore restrict the displayed range to $x\leq0.8$. For the PL wave function, the ratio exhibits a
more pronounced $x$ dependence at larger momentum fractions, whereas the Gaussian wave function
yields a comparatively smoother behavior.

The corresponding Mellin moments are summarized in Table~\ref{tab:mellin_kaon}. As for the pion,
the kaon moments decrease monotonically with increasing $n$ and show only a mild dependence on the
choice of wave function. The moments also decrease slightly with increasing nuclear density. Because
the kaon contains quarks of different masses, namely a light quark $q$ and a strange quark $s$,
their Mellin moments are distinct, reflecting the asymmetric momentum distributions of the two
constituents.

\begin{table}[ht]
\begin{ruledtabular}
		\renewcommand{\arraystretch}{1.2}
\caption{Evolved Mellin moments $\langle x^n\rangle_q$ of the $D^\pm$ PDF obtained with the Gaussian (PL) wave functions at $\mu^2=4~\mathrm{GeV}^2$ for different nuclear densities.}
\label{tab:mellin_D}
\begin{tabular}{c|ccc}
$\langle x^n \rangle_q$ & $\rho/\rho_0 = 0.0$  & $\rho/\rho_0 = 0.5$  & $\rho/\rho_0 = 1.0$  \\
\hline
$\langle x^{1} \rangle_q$ &  0.109(0.093) &  0.105(0.086) &  0.102(0.082)  \\
$\langle x^{2} \rangle_q$ &  0.024(0.017) &  0.023(0.016) &  0.022(0.015)  \\
$\langle x^{3} \rangle_q$ &  0.007(0.004) &  0.007(0.004) &  0.006(0.004)  \\
$\langle x^{4} \rangle_q$ &  0.002(0.001) &  0.002(0.001) &  0.002(0.001)  \\
$\langle x^{5} \rangle_q$ &  0.001(0.001) &  0.001(0.001) &  0.001(0.001)
\end{tabular}
	\end{ruledtabular}
\end{table}

\begin{table}[ht]
\begin{ruledtabular}
    \renewcommand{\arraystretch}{1.2}
\caption{Mellin-moment ratios $R_n=\langle x^n\rangle_q/\langle x^n\rangle_c$ of the $D^\pm$ meson obtained with the Gaussian (PL) wave function.}
\label{tab:mellin_ratios_D}
\begin{tabular}{cccc}
    $n$ & $R_n$ ($\rho = 0.0$) & $R_n$ ($\rho = 0.5$) & $R_n$ ($\rho = 1.0$) \\
    \hline
    1 & 1.004 (1.005) & 1.004 (1.006) & 1.004 (1.006) \\
    2 & 0.471 (0.378) & 0.467 (0.374) & 0.466 (0.374) \\
    3 & 0.252 (0.172) & 0.249 (0.170) & 0.249 (0.171) \\
    4 & 0.146 (0.090) & 0.144 (0.089) & 0.143 (0.090) \\
    5 & 0.089 (0.052) & 0.087 (0.052) & 0.087 (0.053) \\
\end{tabular}
\end{ruledtabular}
\end{table}

For clarity, we further characterize the flavor dependence by considering the ratio of Mellin
moments, $ R_n=\langle x^n\rangle_q/\langle x^n\rangle_{\bar q}$. For the $K^\pm$ meson, we first
evaluate the ratio $\langle x^n\rangle_q/\langle x^n\rangle_s$, with the results summarized in
Table~\ref{tab:mellin_ratios_kaon}. The Mellin moments of the strange quark are found to be larger
than those of the light quark for both wave-function ansätze. Overall, the ratio decreases with
increasing $n$, indicating that the strange-quark contribution becomes increasingly important at
larger momentum fractions, while the relative contribution of the light quark is suppressed. A
similar trend has been reported in Ref.~\cite{Alexandrou:2026nsl}. With increasing nuclear density,
the ratio exhibits a slight increase, except for $n=1$, indicating that the medium modification
affects the relative momentum distributions of the light and strange quarks. In the approximate
SU(3)-flavor-symmetric limit, the first moment ratio $R_1$ is close to unity, although the in-medium
modifications prevent exact equality. More generally, the deviation from SU(3) symmetry is
substantial, reaching approximately $30\%$--$60\%$ depending on the moment and wave-function ansatz,
with the breaking becoming more pronounced for higher moments.

\paragraph{$D$ meson PDFs}
Results for the $D^\pm$ meson PDFs in vacuum and nuclear matter at $\mu^2=4~\mathrm{GeV}^2$ (left
panels) and $27~\mathrm{GeV}^2$ (right panels) are shown in Fig.~\ref{fig15}(a)--(d). For the
$D^\pm$ and $B^\pm$ mesons, the momentum-fraction axis is displayed on a logarithmic scale to better
resolve the partonic structure at small $x$. As shown in Fig.~\ref{fig15}(a)--(b), the in-medium
modification produces a modest enhancement of the $D^\pm$ PDF at small $x$, followed by a turnover.
For the PL wave function, the distribution begins to decrease at approximately $x\simeq0.07$,
whereas the Gaussian wave function exhibits a similar turnover at a smaller momentum fraction,
$x\simeq0.03$.

Furthermore, the PDF ratios $R_{\rho/0}$ for the $D^\pm$ meson, defined as the ratio of the
in-medium to vacuum PDFs, are shown in Fig.~\ref{fig15}(e)--(h). As in the kaon case, the ratios
exhibit qualitatively similar behavior for the two wave-function ansätze and become ill-defined near
$x\simeq0.7$ as the PDFs approach zero. We therefore restrict the displayed range to $x\leq0.6$. In
the intermediate-$x$ region, the Gaussian wave function yields a ratio that approaches approximately
$0.95$, whereas the PL wave function exhibits a stronger $x$ dependence before approaching a similar
value near $x=0.6$.

The corresponding Mellin moments of the $D^\pm$ meson PDF are summarized in
Table~\ref{tab:mellin_D}. As in the pion and kaon cases, the moments decrease with increasing $n$,
reflecting the dominance of the heavy charm quark at larger momentum fractions. With increasing
nuclear density, the moments exhibit a slight decrease, although the effect becomes negligible for
$n\gtrsim4$. We also evaluate the moment ratio $\langle x^n\rangle_q/\langle x^n\rangle_c$, with the
results listed in Table~\ref{tab:mellin_ratios_D}. Similar to the kaon case, this ratio decreases
with increasing $n$, indicating the increasing dominance of the charm-quark contribution to the
higher Mellin moments.

\begin{table}[ht]
\begin{ruledtabular}
		\renewcommand{\arraystretch}{1.2}
\caption{Evolved Mellin moments $\langle x^n\rangle_q$ of the $B^\pm$ PDF obtained with the Gaussian (PL) wave functions at $\mu^2=4~\mathrm{GeV}^2$ for different nuclear densities.}
\label{tab:mellin_B}
\begin{tabular}{c|ccc}
$\langle x^n \rangle_q$ & $\rho/\rho_0 = 0.0$ (G) & $\rho/\rho_0 = 0.5$ (G) & $\rho/\rho_0 = 1.0$
(G) \\
\hline
$\langle x^{1} \rangle_q$ & 0.053(0.041) & 0.051(0.038) & 0.050(0.037)    \\
$\langle x^{2} \rangle_q$ & 0.006(0.004) & 0.006(0.003) & 0.006(0.003)    \\
$\langle x^{3} \rangle_q$ & 0.001($5\times10^{-4}$) & 0.001($5\times10^{-4}$) &
0.001($5\times10^{-4}$)    \\
$\langle x^{4} \rangle_q$ & $2\times10^{-4}$($1\times10^{-4}$) &  $2\times10^{-4}$($1\times10^{-4}$)
& $2\times10^{-4}$($1\times10^{-4}$)  \\
$\langle x^{5} \rangle_q$ & $4\times10^{-5}$($2\times10^{-5}$) &  $3\times10^{-5}$($2\times10^{-5}$)
& $3\times10^{-5}$($2\times10^{-5}$)
\end{tabular}
	\end{ruledtabular}
\end{table}

\begin{table}[ht]
\begin{ruledtabular}
		\renewcommand{\arraystretch}{1.2}
\caption{Mellin-moment ratios $R_n=\langle x^n\rangle_q/\langle x^n\rangle_b$ of the $B^\pm$ meson
obtained with the Gaussian (PL) wave function.}
\label{tab:mellin_ratios_B}
\begin{tabular}{cccc}
    $n$ & $R_n$ ($\rho = 0.0$) & $R_n$ ($\rho = 0.5$) & $R_n$ ($\rho = 1.0$) \\
    \hline
    1 & 1.009 (1.012) & 1.010 (1.012) & 1.010 (1.012) \\
    2 & 0.197 (0.154) & 0.196 (0.154) & 0.196 (0.154) \\
    3 & 0.047 (0.032) & 0.046 (0.032) & 0.046 (0.032) \\
    4 & 0.013 (0.008) & 0.013 (0.009) & 0.013 (0.009) \\
    5 & 0.004 (0.003) & 0.004 (0.003) & 0.004 (0.003) \\
\end{tabular}
    \end{ruledtabular}
\end{table}

\paragraph{$B$ meson PDFs}
Figure~\ref{fig16}(a)--(b) shows the $B^\pm$ meson PDFs obtained with the PL wave function at
$\mu^2=4$ and $27~\mathrm{GeV}^2$, respectively, for several nuclear densities. As for the $D^\pm$
meson, the $x$-axis is displayed on a logarithmic scale to resolve the small-$x$ behavior more
clearly. The $B^\pm$ PDF exhibits a qualitative pattern similar to that of the $D^\pm$ meson:
increasing the nuclear density enhances the distribution over a limited range of $x$, followed by a
decrease near the peak.

The PDF ratios $R_{\rho/0}$ are shown in Fig.~\ref{fig16}(e)--(h). As in the kaon and $D^\pm$
cases, the ratios become ill-defined near $x\simeq0.5$ as the corresponding PDFs approach zero. The
Gaussian wave function exhibits a weaker density dependence, whereas the PL wave function shows a
more pronounced variation in the intermediate-$x$ region. At the two evolution scales 4 and 27
GeV$^2$, the PL ratios approach approximately $0.90$ and $0.96$, respectively.

The corresponding Mellin moments of the $B^\pm$ meson PDF are summarized in
Table~\ref{tab:mellin_B}. As in the $D^\pm$ case, the moments decrease substantially with increasing
$n$, reflecting the dominance of the heavy $b$ quark at large momentum fractions. The density
dependence is relatively weak, becoming negligible for $n\gtrsim3$. We also evaluate the moment
ratio $\langle x^n\rangle_q/\langle x^n\rangle_b$, with the results listed in
Table~\ref{tab:mellin_ratios_B}. In contrast to the $D^\pm$ meson, the ratio exhibits a pronounced
change between $n=1$ and $n=2$ for both wave-function ansätze. It subsequently decreases with
increasing $n$, reflecting the increasingly dominant contribution of the $b$ quark to the higher
Mellin moments.

\section{Summary}
\label{sec:sum}
In summary, we have systematically investigated the DAs and PDFs of the pion, kaon, and heavy-light
pseudoscalar mesons in nuclear matter within the LFQM-QMC framework. We considered both Gaussian and
PL light-front wave functions and evolved the distributions to higher scales using the ERBL and
DGLAP evolution equations. For the DAs, we analyzed the corresponding $\xi$ moments and Gegenbauer
coefficients, while for the PDFs, we examined the Mellin moments and in-medium-to-vacuum ratios.

Our results show that the Gaussian and PL wave functions yield qualitatively similar LFWFs, DAs,
and PDFs in both vacuum and nuclear matter, while their distinct large-momentum behavior leads to
quantitative differences, particularly near the endpoints. The vacuum pion and kaon DAs and PDFs are
consistent with available lattice QCD calculations and experimental data, providing a useful
benchmark for the present framework. In nuclear matter, the light pseudoscalar mesons exhibit more
pronounced medium modifications, whereas these effects are progressively suppressed in the
heavy-light $D^\pm$ and $B^\pm$ mesons. QCD evolution to higher scales generally reduces the
sensitivity of the distributions to the nuclear environment while preserving their main qualitative
features.

In nuclear matter, modifications of the in-medium quark properties enhance the LFWFs near the
endpoints while suppressing them in the intermediate-$x$ region. This redistribution is reflected in
the DAs and PDFs, which exhibit a corresponding suppression at intermediate $x$, with QCD evolution
shifting the location of this feature. These results provide a possible connection between medium
modifications of hadron structure and the EMC effect. To quantify the medium dependence, we also
examine the PDF ratios $R_{\rho/0}$. Although their detailed behavior depends on the evolution scale
and wave-function ansatz, the Gaussian and PL wave functions yield consistent qualitative trends.

The present study provides predictions for the DAs and PDFs of pseudoscalar mesons in vacuum and
nuclear matter. In vacuum, these results can be tested against available lattice QCD calculations
and experimental data. Future measurements at the EIC~\cite{Arrington:2021biu},
EicC~\cite{Anderle:2021wcy}, J-PARC~\cite{Sawada:2016mao}, and AMBER/COMPASS++~\cite{Adams:2018pwt}
will provide additional constraints on the free-space distributions, particularly in the small- and
large-$x$ regions. Direct measurements of in-medium meson distributions, however, remain
experimentally challenging. A promising direction for future work is therefore the development of
more self-consistent and systematically controlled frameworks that describe meson structure in both
vacuum and nuclear matter, thereby providing deeper insight into the interplay between hadron
structure and nuclear many-body dynamics.

\section*{Acknowledgement}
M.R. thanks Fernando Chandra (University of Regina, Canada) for useful discussions on PDF QCD evolution and also Craig McNeile for some useful suggestions.
M.R. was supported by the PhD studentship jointly funded by the Science and Technology Facilities
Council (STFC), part of UK Research and Innovation (UKRI), and the School of Engineering, Computing,
and Mathematics at the University of Plymouth.
A.J.A. was supported by the JAEA Postdoctoral Fellowship Program and by the PUTI Q1 Grant from the
University of Indonesia under Contract No. PKS-206/UN2.RST/HKP.05.00/2025. A.J.A, P.T.P.H., and
K.T. were partially supported by the RCNP Collaboration Research Network Program under
Project No.COREnet 057.
K.T. was supported by the National Council for Scientific and Technological Development –
CNPq, Brazil, Processes No. 304199/2022-2 and No. 306866/2026-9, by the São Paulo Research
Foundation (FAPESP), Process No. 2023/07313-6 and No. 2026/01656-7, and by the Instituto Na-
cional de Ciência e Tecnologia - Nuclear Physics and Applications (INCT-FNA), Brazil, Process
No. 408419/2024-5.
This work was supported by the World Premier International Research Center Initiative
(WPI-SKCM$^2$) of Hiroshima University, MEXT, Japan.

\appendix

\begin{table}[ht!]
\begin{ruledtabular}
		\renewcommand{\arraystretch}{1.2}
\caption{Evolved $\xi$ moments of the $\pi^\pm$ DA obtained with the Gaussian and PL wave functions
at $\mu^2=4~\mathrm{GeV}^2$ for different nuclear densities. The odd $\xi$ moments vanish because
of the symmetric quark content of the pion. Lattice QCD results from Ref.~\cite{Cloet:2024vbv} are
included for comparison, with statistical and systematic uncertainties of $\pm0.030$ and $\pm0.020$,
respectively.}
\label{tab:xi_moments_pi}
\begin{tabular}{c|ccc|c}
$\langle \xi^n \rangle$ & $\rho/\rho_0 = 0.0$ (G) & $\rho/\rho_0 = 0.5$ (G) & $\rho/\rho_0 = 1.0$
(G) & ~\cite{Cloet:2024vbv} \\
\hline
$\langle \xi^{2} \rangle$ &  0.233(0.244) &  0.258(0.264) &  0.280(0.282) & 0.267(11) \\
$\langle \xi^{4} \rangle$ &  0.107(0.118) &  0.128(0.134) &  0.149(0.151) & 0.139(10) 
\end{tabular}
	\end{ruledtabular}
\end{table}

\begin{table}[ht!]
\begin{ruledtabular}
		\renewcommand{\arraystretch}{1.2}
\caption{Evolved $\xi$ moments of the $K^\pm$ DA obtained with the Gaussian (G) and PL wave
functions at $\mu^2=4~\mathrm{GeV}^2$ for different nuclear densities. Lattice QCD results from
Ref.~\cite{Cloet:2024vbv} are included for comparison.}
\label{tab:xi_moments_K}
\begin{tabular}{c|ccc|c}
$\langle \xi^n \rangle$ & $\rho/\rho_0 = 0.0$ (G) & $\rho/\rho_0 = 0.5$ (G) & $\rho/\rho_0 = 1.0$
(G) & \cite{Cloet:2024vbv}\\
\hline
$\langle \xi^{1} \rangle$ &  $\phantom{-}$0.012($-0.001$) &  $\phantom{-}$0.002($-0.016$) &
$-0.029(-0.052)$ & $0.001(10)$ \\
$\langle \xi^{2} \rangle$ &  $\phantom{-}$0.224(0.239) &  $\phantom{-}$0.231(0.246) &
$\phantom{-}$0.231(0.246) & $0.237(7)$\\
$\langle \xi^{3} \rangle$ &  $\phantom{-}$0.010(0.002) &  $\phantom{-}$0.010($-0.003$) &
$-0.003(-0.019)$ & - \\
$\langle \xi^{4} \rangle$ &  $\phantom{-}$0.101(0.114) &  $\phantom{-}$0.107(0.119) &
$\phantom{-}$0.107(0.120) & $0.115(6)$ \\
$\langle \xi^{5} \rangle$ &  $\phantom{-}$0.008(0.002) &  $\phantom{-}$0.010(0.0002) &
$\phantom{-}$0.004($-0.009$) &  -
\end{tabular}
	\end{ruledtabular}
\end{table}

\section{$\xi$-moments of in-medium DAs}
\label{app:xi-moment}
In this appendix, we summarize the density dependence and scale evolution of the $\xi$ moments of
the light ($\pi^\pm$, $K^\pm$) and heavy-light ($D^\pm$, $B^\pm$) meson DAs up to $n=6$ at the
renormalization scale $\mu^2=4~\mathrm{GeV}^2$.

\paragraph{Pion DAs} The evolved $\xi$ moments of the pion DA at $\mu^2=4~\mathrm{GeV}^2$ up to
$n=6$ are summarized in Table~\ref{tab:xi_moments_pi}. The moments generally increase with
increasing nuclear density. Owing to isospin and charge-conjugation symmetry, all odd $\xi$ moments
of the pion DA vanish. In vacuum, we obtain $\langle \xi^2\rangle_\pi=0.233$ and $0.244$ for the
Gaussian and PL wave functions, respectively. These values are consistent with lattice QCD
determinations at $\mu^2=4~\mathrm{GeV}^2$. In particular, Ref.~\cite{Zhang:2020gaj} reports
$\langle \xi^2\rangle_\pi=0.244(30)(20)$, while Ref.~\cite{Braun:2015axa} finds $\langle
\xi^2\rangle_\pi=0.236(41)(39)$. Our vacuum predictions are also consistent with the other
lattice QCD determinations~\cite{Braun:2006dg,Arthur:2010xf,Bali:2017ude,RQCD:2019osh,Cloet:2024vbv,Detmold:2025lyb}.

\paragraph{Kaon DAs}The $\xi$ moments of the kaon DA obtained with the Gaussian and PL wave
functions at $\mu^2=4~\mathrm{GeV}^2$ are listed in Table~\ref{tab:xi_moments_K}. In contrast to the
pion, the odd moments of the kaon DA are nonzero, reflecting SU(3)-flavor symmetry breaking between
the light and strange quarks. As the nuclear density increases, the odd $\xi$ moments decrease,
whereas the even moments increase.

At $\mu^2=4~\mathrm{GeV}^2$, our results are consistent with available lattice QCD determinations.
In particular, the RQCD result $\langle\xi^2\rangle_K=0.231^{+4}_{-4}$~\cite{RQCD:2019osh} agrees
with both our Gaussian and PL wave-function results. For the Gaussian wave function in vacuum, we
obtain $\langle\xi\rangle_K=0.012$ and $\langle\xi^2\rangle_K=0.224$, which are compatible with the
lattice results of Ref.~\cite{Zhang:2020gaj}, $\langle\xi\rangle_K=0.007(16)$--$0.009(18)$ and
$\langle\xi^2\rangle_K=0.194(14)(16)$--$0.198(16)(16)$, as well as those of
Ref.~\cite{Braun:2006dg}, $\langle\xi\rangle_K=0.0272(5)(17)$ and $\langle\xi^2\rangle_K=0.260(6)$.

\begin{table}[ht!]
\begin{ruledtabular}
		\renewcommand{\arraystretch}{1.2}
\caption{Evolved $\xi$ moments of the $D^\pm$ meson DA obtained with the Gaussian (G) and PL wave
functions at $\mu^2=4~\mathrm{GeV}^2$ for different nuclear densities. QCD sum-rule results from
Ref.~\cite{Zhong:2022ugk} are included for comparison.}
\label{tab:xi_moments_D}
\begin{tabular}{c|ccc|c}
$\langle \xi^n \rangle$ & $\rho/\rho_0 = 0.0$ & $\rho/\rho_0 = 0.5$ & $\rho/\rho_0 = 1.0$ &
~\cite{Zhong:2022ugk} \\
\hline
$\langle \xi^{1} \rangle$ & $-0.308(-0.294)$ & $-0.304(-0.292)$ & $-0.294(-0.283)$ & $-0.349(37)$ \\
$\langle \xi^{2} \rangle$ &  $\phantom{-}$0.230(0.253) &  $\phantom{-}$0.229(0.256) &
$\phantom{-}$0.222(0.252) & $\phantom{-}$0.251(14)\\
$\langle \xi^{3} \rangle$ & $-0.132(-0.132)$ & $-0.132(-0.135)$ & $-0.125(-0.130)$ & $-0.152(12)$ \\
$\langle \xi^{4} \rangle$ &  $\phantom{-}$0.109(0.121) &  $\phantom{-}$0.109(0.125) &
$\phantom{-}$0.104(0.123) & - \\
$\langle \xi^{5} \rangle$ & $-0.074(-0.076)$ & $-0.075(-0.079)$ & $-0.071(-0.077)$ & - 
\end{tabular}
\renewcommand{\arraystretch}{1}
	\end{ruledtabular}
\end{table}

\begin{table}[ht!]
\begin{ruledtabular}
\renewcommand{\arraystretch}{1.2}
\caption{Evolved $\xi$ moments of the $B^\pm$ meson DA obtained with the Gaussian and PL wave
functions at $\mu^2=4~\mathrm{GeV}^2$ for different nuclear densities.}
\label{tab:xi_moments_B}
\begin{tabular}{c|ccc}
$\langle \xi^n \rangle$ & $\rho/\rho_0 = 0.0$ & $\rho/\rho_0 = 0.5$ & $\rho/\rho_0 = 1.0$ \\
\hline
$\langle \xi^{1} \rangle$ & $-0.553(-0.549)$ & $-0.551(-0.548)$ & $-0.546(-0.543)$ \\
$\langle \xi^{2} \rangle$ &  $\phantom{-}$0.402(0.434) &  $\phantom{-}$0.400(0.435) &
$\phantom{-}$0.394(0.431) \\
$\langle \xi^{3} \rangle$ & $-0.285(-0.319)$ & $-0.283(-0.321)$ & $-0.277(-0.317)$ \\
$\langle \xi^{4} \rangle$ &  $\phantom{-}$0.227(0.267) &  $\phantom{-}$0.225(0.270) &
$\phantom{-}$0.219(0.267) \\
$\langle \xi^{5} \rangle$ & $-0.176(-0.213)$ & $-0.174(-0.217)$ & $-0.169(-0.214)$ 
\end{tabular}
\renewcommand{\arraystretch}{1.2}
	\end{ruledtabular}
\end{table}

For the PL wave function, we find $\langle\xi\rangle_K=-0.001$, which has the opposite sign to the
lattice QCD results of Refs.~\cite{Zhang:2020gaj,Braun:2006dg}. A negative first moment, however,
has also been reported for the kaon DA in Ref.~\cite{Nam:2006au}.

\paragraph{Heavy meson DAs} The $\xi$ moments of the heavy-light meson DAs exhibit distinct density
dependences that vary with the moment order. For the $D^\pm$ meson at $\mu^2=4~\mathrm{GeV}^2$, as
shown in Table~\ref{tab:xi_moments_D}, the even moments increase as the density changes from
$\rho/\rho_0=0$ to $0.5$ but decrease at $\rho/\rho_0=1$. The first and third odd moments increase
with density while remaining negative throughout the range $\rho/\rho_0=0$, $0.5$, and $1$. The
fifth moment shows a nonmonotonic density dependence, decreasing at $\rho/\rho_0=0.5$ and increasing
again at $\rho/\rho_0=1$ relative to its vacuum value, while remaining negative.

For the $B^\pm$ meson at $\mu^2=4~\mathrm{GeV}^2$, the results in Table~\ref{tab:xi_moments_B} show
that all even $\xi$ moments are positive, whereas all odd moments are negative. In vacuum, both the
overall shape of the $B^\pm$ DA and the signs of the lowest moments, $\langle\xi\rangle$ and
$\langle\xi^2\rangle$, are in good agreement with the lattice QCD results of
Ref.~\cite{Braun:2003wx}.

\bibliography{references}

\end{document}